\documentclass{aa}  

\usepackage{graphicx}
\usepackage[table]{xcolor}
\usepackage{txfonts}
\usepackage{bm}
\usepackage{multirow, makecell}
\usepackage{booktabs}
\usepackage{fontawesome}
\newcommand{\github}{\href{https://github.com/dlanzieri/WL_Implicit-Inference}{\faGithub}} %For Denise: Remember to label the final version of the Github repo with the notebook
\usepackage[colorlinks=true,citecolor=blue,linkcolor=blue,urlcolor=blue]{hyperref}
\begin{document} 

\newcommand{\sbilens}{\texttt{sbi\_lens}}

\title{Optimal Neural Summarisation for Full-Field Weak Lensing Cosmological Implicit Inference}

\definecolor{mypink}{HTML}{e0218a}

\newcommand{\justine}[1]{{\color{cyan}JZ: #1}}
\newcommand{\reformulate}[1]{{\color{mypink}remove: #1}}
\newcommand{\denise}[1]{{\color{red}DL: #1}}
\newcommand{\FL}[1]{{\color{magenta}FL: #1}}
\newcommand{\LM}[1]{{\color{olive}LM: #1}}

\author{Denise Lanzieri \inst{1,}\inst{5} \thanks{equal contribution} 
\and
Justine Zeghal \inst{2,8,9,10 \: \star}
\and
T. Lucas Makinen \inst{3}
\and
Alexandre Boucaud \inst{2}
\and
Jean-Luc Starck \inst{4, } \inst{6}
\and
Fran\c{c}ois Lanusse \inst{4,7}
}
\institute{Université Paris Cité, Université Paris-Saclay, CEA, CNRS, AIM, F-91191, Gif-sur-Yvette, France
\and
Université Paris Cité, CNRS, Astroparticule et Cosmologie, F-75013 Paris, France
\and Imperial Centre for Inference and Cosmology (ICIC) $\&$ Astrophysics Group, Imperial College London, Blackett Laboratory, Prince Consort Road, London SW7 2AZ, United Kingdom
\and
Université Paris-Saclay, Université Paris Cité, CEA, CNRS, AIM, 91191, Gif-sur-Yvette, France
\and 
 Sony Computer Science Laboratories - Rome, Joint Initiative CREF-SONY, Centro Ricerche Enrico Fermi, Via Panisperna 89/A, 00184, Rome, Italy
 \and 
 Institutes of Computer Science and Astrophysics, Foundation for Research and Technology Hellas (FORTH), Greece
 \and Center for Computational Astrophysics, Flatiron Institute, 162 5th Ave, New York, NY, 10010, USA
 \and Department of Physics, Université de Montréal, Montréal H2V 0B3, Canada
 \and Mila – Quebec Artificial Intelligence Institute, Montréal H2S 3H1, Canada
 \and Ciela – Montreal Institute for Astrophysical Data Analysis and Machine Learning, Montréal H2V 0B3, Canada
}
\titlerunning{}
\date{Received xxx; accepted xxx}

  \abstract
   { Traditionally, weak lensing cosmological surveys have been analyzed using summary statistics that were either motivated by their analytically tractable likelihoods (e.g. power spectrum), or by their ability to access \textit{some} higher-order information (e.g. peak counts) but in that case at the cost of requiring a Simulation-Based Inference approach. In both cases, even if these statistics can be very informative, they are not designed nor guaranteed to be statistically sufficient (i.e. to capture all of the cosmological information content of the data). With the rise of deep learning, however, it becomes possible to create summary statistics that are specifically optimized to extract the full cosmological information content of the data. However, a fairly wide range of loss functions have been used in practice in the weak lensing literature to train such neural networks, leading to the natural question of whether a given loss should be preferred and whether sufficient statistics can be achieved in theory and in practice under these different choices.
   }   
    { We compare different neural summarization strategies that have been proposed in the literature, to answer the question of what loss function leads to theoretically optimal summary statistics for performing full-field cosmological inference. In doing so, we aim to provide guidelines and insights to the community to help guide future neural network-based cosmological inference analyses.  
    }  
    { We design an experimental setup which allows us to isolate the specific impact of the loss function used to train neural summary statistics on weak lensing data, at fixed neural architecture and Simulation-Based Inference pipeline. To achieve this, we have developed the \href{https://github.com/DifferentiableUniverseInitiative/sbi_lens}{\sbilens} JAX package, which implements an automatically differentiable lognormal weak lensing simulator and the tools needed to perform explicit full-field inference with a Hamiltonian-Monte Carlo (HMC) sampler over this model. Using \sbilens, we simulate a $w$CDM LSST Year 10 weak lensing analysis scenario in which the full-field posterior obtained by HMC sampling gives us a ground truth to which to compare different neural summarization strategies.     
    } 
   { We provide theoretical insight into the different loss functions being used in the literature (e.g. Mean Square Error (MSE) regression) and show that some do not necessarily lead to sufficient statistics, while those motivated by information theory (in particular Variational Mutual Information Maximization (VMIM)) can in principle lead to optimal sufficient statistics. Our numerical experiments confirm these insights and we show in particular in our simulated $w$CDM scenario that the Figure of Merit (FoM) of an analysis using a neural summary statistics optimized under VMIM achieves 100\% of the reference $\Omega_c - \sigma_8$ full-field FoM, while an analysis using a summary statistics trained under simple MSE achieves only 81\% of the same reference FoM. To facilitate further experimentation and benchmarking by the community we make our simulation framework  \href{https://github.com/DifferentiableUniverseInitiative/sbi_lens}{\sbilens} publicly available, and all codes associated with the analyses presented in this paper are available at this link \github \hspace{0.2mm}.}
   {}

   \maketitle

%-------------------------------------------------------------------
%------------------------------------------------------------------
%------------------------------------------------------------------
%------------------------------------------------------------------
\section{Introduction}
Weak gravitational lensing by large-scale structures (LSS) is caused by the presence of foreground matter bending the light emitted by background galaxies. Being sensitive to the large-scale structure of the universe, it is one of the most promising tools for investigating the nature of dark energy, the origin of the accelerating expansion of the universe, and estimating cosmological parameters. Future cosmological surveys, such as the Legacy Survey of Space and Time (LSST) of the Vera C. Rubin Observatory \citep{ivezic2019lsst}, Nancy Grace Roman Space Telescope \citep{spergel2015wide}, and the Euclid Mission \citep{laureijs2011euclid}, will rely on weak gravitational lensing as one of the principal physical probes to address unresolved questions in current cosmology.
 As these surveys become deeper, they will be able to access more non-Gaussian features of the matter fields. This makes the standard weak-lensing analyses, which rely on two-point statistics such as the 2-point shear correlation or the angular power spectrum, sub-optimal. These analyses are unable to fully capture the non-Gaussian information imprinted in the lensing signal and can only access Gaussian information.

To overcome this limitation, several higher-order statistics have been introduced. These include weak-lensing peak counts \citep{liu2015cosmology,  liu2015cosmological, lin2015new, kacprzak2016cosmology, peel2017cosmological, shan2018kids, martinet2018kids, ajani2020constraining, harnois2021cosmic, zurcher2022dark}, wavelet and scattering transform \citep{ajani2021starlet, cheng2021weak}, the one-point probability distribution function \citep[PDF][]{liu2019constraining, uhlemann2020fisher, boyle2021nuw}, Minkowski functionals \citep{kratochvil2012probing, petri2013cosmology}, moments of mass maps \citep{gatti2021dark},  and three-point  point statistics \citep{takada2004cosmological, semboloni2011weak, rizzato2019tomographic, halder2021integrated}. 
Although these methods have proven to provide additional cosmological information beyond the power spectrum, they do not necessarily exhaust the information content of the data.

In recent years, the idea of performing full-field inference (i.e. to analyze the full information content of the data at the cosmological field level) has gained traction and contrary to previous approaches directly aims by design at achieving information-theoretically optimal posterior contours.

Within this category, a further distinction can be made between \textit{explicit} methods that rely on a Bayesian Hierarchical Model (BHM) describing the joint likelihood $p(\bm{x}|\bm{\theta},\bm{z})$ of the field-level data $\bm{x}$, cosmological parameters $\bm{\theta}$, and as well as all latent parameters $\bm{z}$ of the model \citep[e.g.][]{alsing2017cosmological, porqueres2021bayesian, porqueres2023field, Karma2022, Fiedorowicz_2022, Zhou2023}, and \textit{implicit} methods (a.k.a Simulation-Based Inference, or Likelihood-Free Inference) that rely on neural networks to directly estimate the full-field marginal likelihood $p(\bm{x}|\bm{\theta})$ or posterior  $p(\bm{\theta}|\bm{x})$ on cosmological parameters from a black-box simulation model \citep{2018PhRvD..97j3515G, fluri2018cosmological, fluri2019cosmological, ribli2019weak, PhysRevD.102.123506, jeffrey2021likelihood, fluri2021cosmological, fluri2022full, lu2022simultaneously, kacprzak2022deeplss, lu2023cosmological, Akhmetzhanova2024, Jeffrey2024}. Finally, another recently proposed flavor of implicit methods relies on directly estimating the full-field likelihood \citep{Dai2024}, in contrast to the previous methods which only target a marginal likelihood.

Despite the differences among these works in terms of the physical models they assume or the methodology employed, they all demonstrate that conducting a field-level analysis results in more precise constraints compared to a conventional two-point function analysis. However open questions and challenges remain with both categories of methods. Explicit inference methods have not yet been successfully applied to data in order to constrain cosmology, and are particularly challenging to scale to the volume and resolution of modern surveys. Implicit inference methods have proven to be much easier to deploy in practice and are already leading to state-of-the-art cosmological results \citep{Jeffrey2024, lu2023cosmological, fluri2022full}, but the community has not yet fully converged on a set of best practices for this approach, which transpires in the variety of strategies found in the literature.

In this paper, we aim to clarify part of the design space of implicit inference methods related to how to design optimal data compression procedures, aiming to derive low-dimensional summary statistics while minimizing information loss. Such neural summarization of the data is typically the first step in a two-step strategy found in most implicit inference work: 1) neural compression of the data, in which a neural network is trained on simulations to compress shear or convergence maps down to low-dimensional summary statistics. 2) density estimation, in which either the likelihood of these summary statistics or directly the posterior distribution is estimated from simulations. 

To quantify the extraction power of neural compression schemes found in the literature, we design an experimental setup that allows us to isolate the impact of the loss function used to train the neural network. Specifically, we have developed the \sbilens\ package which provides a differentiable lognormal forward model from which we can simulate a $w$CDM LSST Year 10 weak lensing analysis scenario. The differentiability of our forward model enables us to conduct an explicit full-field inference through the use of Hamiltonian Monte Carlo (HMC) sampling scheme. This explicit posterior serves as our ground truth against which we quantify the quality of the benchmarked compression procedures according to sufficiency definition. To only assess the impact of the loss function of the compression step, we fix both the inference methodology and the neural compressor architecture. In addition to empirical results, we provide some theoretical insight into the different losses employed in the literature. We explain why Mean Square Error (MSE), Mean Absolute Error (MAE) or Gaussian Negative Log Likelihood (GNLL) losses might lead to sub-optimal summary statistics while information theory motivated losses such as Variational Mutual Information Maximization (VMIM) theoretically achieve sufficient statistics.

The paper is structured as follows: in Section \ref{Sec:Motivation}, we illustrate the motivation behind this work. In Section \ref{sec:compression_review} we provide a detailed theoretical overview of the loss functions commonly used for compression. In Section \ref{Sec:the SBILens framework}, we introduce the \sbilens\ framework and describe the simulated data used in this work. In Section \ref{Sec:experiment}, we detail the inference strategy and the three different approaches we used: the power spectrum, explicit full-field inference and implicit full-field inference. In Section \ref{Sec:results}, we discuss the results and validate the implicit inference approaches. Finally, we conclude in Section \ref{Sec:conclusion}. 
%------------------------------------------------------------------
%------------------------------------------------------------------
%------------------------------------------------------------------
%------------------------------------------------------------------
%--------------------------------------------------------------------
%--------------------------------------------------------------------
%--------------------------------------------------------------------
% ############# Begin Table BiblioSruvey  #############
%--------------------------------------------------------------------
\begin{table*}
    \begin{center}
        \begin{tabular}{ |p{5cm}|p{3cm}|p{5cm}| }
            \hline
            \makecell{\textbf{Reference}} &  \makecell{\textbf{Loss function}} &  \makecell{\textbf{Inference strategy}}  \\
            \hline
            \citet{2018PhRvD..97j3515G} & \makecell{MAE} &  \makecell{Likelihood-based analysis} \\
            \hline
            \citet{fluri2018cosmological} & \makecell{GNLL} & \makecell{Likelihood-based analysis}
            \\
            \hline     
             \rowcolor{lightgray}
                        \citet{fluri2019cosmological} &  \makecell{GNLL} &  \makecell{Likelihood-based analysis}
            \\
            \hline                    
                        \citet{ribli2019weak} & \makecell{MAE} & \makecell{Likelihood-based analysis}
            \\            
            \hline             
                        \citet{PhysRevD.102.123506} & \makecell{MAE} & \makecell{Likelihood-based analysis}
            \\
            \hline 
             \rowcolor{lightgray}
                        \citet{jeffrey2021likelihood} & 
                        \makecell{MSE \\ VMIM}
                        &  \makecell{Likelihood Free Inference \\ (Py-Delfi)} 
            \\            
            \hline             
                        \citet{fluri2021cosmological} &\makecell{IMNN} &   \makecell{Likelihood Free Inference \\ (GPABC)}  
            \\
            \hline
             \rowcolor{lightgray}
                        \citet{fluri2022full} &  \makecell{IMNN} & \makecell{Likelihood Free Inference \\ (GPABC)}
            \\            
            \hline 
                        \citet{lu2022simultaneously} & \makecell{MSE} & \makecell{Likelihood-based analysis}   
            \\
            \hline 
                        \citet{kacprzak2022deeplss} & \makecell{GNLL}  & \makecell{Likelihood-based analysis}
            \\
            \hline 
             \rowcolor{lightgray}
                        \citet{lu2023cosmological} &  \makecell{MSE} & \makecell{Likelihood-based analysis} 
            \\
            \hline 
                        \citet{Akhmetzhanova2024} &  \makecell{VICReg} & \makecell{Likelihood Free Inference  \\(SNPE)} 
            \\
            \hline 
                        \citet{Charma2024} &  \makecell{MSE, MSE\textsubscript{PCA}, \\ MSE\textsubscript{NP}, VMIM} & \makecell{Likelihood-based analysis} \\
            \hline 
             \rowcolor{lightgray}
                        \citet{Jeffrey2024} &  \makecell{MSE} & \makecell{Likelihood Free Inference \\(Py-Delfi)} \\
                        
            \hline      
        \end{tabular}
        \caption{Table summarizing the different neural compression schemes used for weak-lensing applications. Gray boxes correspond to analyses performed on real data. 
        \\
        Abbreviations used in the Table: MSE-Mean Squared Error; MSE\textsubscript{NP}-Mean Squared Error in $S_8$ space; MSE\textsubscript{PCA}-Mean Squared Error in PCA space; MAE-Mean Absolute Error; GNLL- Gaussian Negative Log Likelihood; VMIM- Variational Mutual Information Maximization; VICReg: Variance-Invariance-Covariance Regularization; IMNN- Information Maximizing Neural Networks; GPABC-Gaussian Processes Approximate Bayesian Computation.}
        \label{tab:biblio_survey}
    \end{center}
\end{table*}
%--------------------------------------------------------------------
% ############# End Table BiblioSruvey  #############
%--------------------------------------------------------------------
\section{Motivation}\label{Sec:Motivation}
%--------------------------------------------------------------------
With the increased statistical power of stage IV surveys, conventional summary statistics such as the power spectrum but also higher-order statistics like peak counts may not fully capture the non-Gaussian information present in the lensing field at the scales accessible to future surveys. In this paper, we focus on full-field inference methods which aim to preserve all available information and facilitate the incorporation of systematic effects and the combination of multiple cosmological probes through joint simulations. \\

In a forward model context, the simulator of the observables serves as our physical model. These models, often referred to as \textit{probabilistic program}, as illustrated by \citet{cranmer2020frontier}, can be described as follows: the models take as input a vector parameter $\bm{\theta}$. Then, they sample internal states $\bm{z}$, dubbed \textit{latent variables}, from the distribution $p(\bm{z}|\bm \theta)$. These states can be directly or indirectly related to a physically meaningful state of the system. Finally, the models generate the output $\bm x$ from the distribution $p(\bm x|\bm \theta,\bm{z})$, where $\bm{x}$ represents the observations.
    
The ultimate goal of Bayesian inference in cosmology is to compute the posterior distribution:
\begin{equation}\label{Eq:posterior}
     p(\bm{\theta}|\bm{x})= 
     \frac{p(\bm{x}|\bm{\theta})p(\bm{\theta})}
     {\int d\bm{\theta'}p(\bm{x}|\bm{\theta}')p(\bm{\theta}')},
\end{equation}
however, a problems arises because the marginal likelihood $p(\bm{x}|\bm{\theta})$ is typically intractable:
\begin{equation}
    p(\bm{x}|\bm{\theta})=\int p(\bm{x},\bm{z}|\bm{\theta}) d\bm{z}=\int p(\bm{x}|\bm{z},\bm{\theta})p(\bm{z}|\bm{\theta}) d\bm{z},
\end{equation}
since it involves integrating over all potential paths through the latent space.
To overcome this limitation while still capturing the full information content of the data, two different approaches have been proposed in the literature. Although these approaches are often referred to by different names, hereinafter we will make the following distinction:
%--------------------------------------------------------------------
\paragraph{\textbf{Explicit inference}} referring to all likelihood-based inference approaches. 

In the context of full-field inference, this approach can be used when the simulator is built as a tractable probabilistic model. This probabilistic model provides a likelihood $p(\bm{x}|\bm{z},\bm{\theta})$ that can be evaluated. Hence, the joint posterior
\begin{equation}
        p(\bm{\theta},\bm{z}|\bm{x})\propto  p(\bm{x}|\bm{z},\bm{\theta}) p(\bm{z}|\bm{\theta})p(\bm{\theta}),
\end{equation}
can be sampled through Markov Chain Monte Carlo (MCMC) schemes.  
In other words, this approach involves using a synthetic physical model to predict observations and then comparing these predictions with real observations to infer the parameters of the model.

%--------------------------------------------------------------------
\paragraph{\textbf{Implicit inference}} referring to the approaches that infer the distributions (posterior, likelihood, or likelihood ratio) from simulations only. This second class of approaches can be used when the simulator is a black box with only the ability to sample from the joint distribution:
\begin{equation}
    (\bm{x}, \bm{\theta})\sim p(\bm{x}, \bm{\theta}).
\end{equation}
Within this class of methods, we can differentiate between traditional methods such as Approximate Bayesian Computation (ABC) and neural-based density estimation methods. ABC, in its simplest form, employs rejection-criteria-based approaches to approximate the likelihood by comparing simulations with the observation. 
In this work, our focus will be on the second class of methods. 

The standard deep learning based approach for implicit inference can be described as two distinct steps:
\begin{enumerate}
    \item learning an optimal low-dimensional set of summary statistics;
    \item using Neural Density Estimator (NDE) in low dimensions to infer the target distributions.
\end{enumerate}

In the first step, we introduce a parametric function $F_{\varphi}$ such that:
    \begin{equation}
         \bm{t}=F_{\varphi}(\bm{x}),
    \end{equation}
which aims at reducing the dimensionality of the data while preserving information, or, in other words, aims at summarizing the data $\bm{x}$ into sufficient statistics $\bm{t}$. 
By definition \citep{bayesiantheory}, a statistic $\bm{t}$ is said to be sufficient for the parameters $\bm{\theta}$ if $
    p(\bm{\theta} \: | \: \bm{x}) = p(\bm{\theta} \: | \: \bm{t}).
$
Meaning that the full data $\bm{x}$ and the summary $\bm{t}$ lead to the same posterior.
Typically, the sufficient statistics $\bm{t}$ are assumed to have the same dimension as $\bm{\theta}$. 

In the second step, NDE can either target building an estimate $p_{\varphi}$ of the likelihood function $p(\bm{x}|\bm{\theta})$ (referred to as Neural Likelihood Estimation (NLE, \citealp{nle1,nle2})), targeting the posterior distribution $p(\bm{\theta}|\bm{x})$, (known as Neural Posterior Estimation (NPE, \citealp{npe1,npe2,npe3})), or the likelihood ratio $r(\theta, x) = p(x \: | \:\theta) \: / \: p(x)$ (Neural Ratio Estimation (NRE, \citealp{lr2,lr1,lr3})). \\

The main motivation of this work is to evaluate the impact of compression strategies on the posterior distribution and determine the ones that provide sufficient statistics. It is important to consider that different neural compression techniques may not lead to the same posterior distribution. Indeed, according to the definition, only a compression scheme that builds sufficient statistics will lead to the true posterior distribution. We summarize the various neural compression strategies found in the literature in \autoref{tab:biblio_survey}. 
Many of these papers have used neural compression techniques that rely on optimizing the Mean Squared Error (MSE) or the Mean Absolute Error (MAE). 
As we will demonstrate in the following sections, this corresponds to training the model to respectively estimate the mean and the median of the posterior distribution. Other papers rely on assuming proxy Gaussian likelihoods and estimate the mean and covariance of these likelihoods from simulations.
Such compression of summaries could be sub-optimal in certain applications, resulting in a loss of information.

\vspace{1cm}
Ultimately, we should keep in mind that, given a simulation model, if a set of sufficient statistics is used, the two approaches should converge to the same posterior. Therefore, the goals of this work will be to: 
\begin{enumerate}
    \item Find an optimal compression strategy for implicit inference techniques.
    \item Showing that by using this optimal compression strategy, both the implicit and explicit full-field methods yield comparable results.
\end{enumerate} 
%------------------------------------------------------------------
%------------------------------------------------------------------
%------------------------------------------------------------------
%------------------------------------------------------------------

\section{Review of common compression loss functions}
\label{sec:compression_review}

To ensure the robustness of implicit inference techniques in cases where forward simulations are high dimensional, it becomes necessary to employ compression techniques that reduce the dimensionality of the data space into summary statistics. 
Specifically, we try to find a function $\bm{t}=F(\bm{x})$, where $\bm{t}$ represents low-dimensional summaries of the original data vector $\bm{x}$. The objective is to build a compression function $F(\bm{x})$ that captures all the information about $\bm{\theta}$ contained in the data $\bm{x}$ while reducing dimensionality. Previous studies (e.g., \citealp{heavens2000massive, Alsing_2018}) have demonstrated that a compression scheme can be achieved where the dimension of the summaries dim($\bm{t}$) is equal to the dimension of the unknown parameters dim($\bm{\theta}$) without any loss of information at the Fisher level. Although these proofs rely on local assumptions, the guiding principle is that by matching the dimensions of $\bm{t}$ and $\bm{\theta}$, we ensure that $\bm{t}$ has just the right "space" to capture the variation in $\bm{\theta}$ embedded in $\bm{x}$.
Furthermore, regression loss functions require that the summary statistics share the same dimensionality as the parameters $\bm{\theta}$. Thus, to best isolate the influence of the loss function on the construction of summary statistics, we use this dimensional convention for all benchmarked compression schemes.

There exist multiple approaches to tackling this dimensionality reduction challenge. This section aims to provide an overview of the various neural compression-based methods employed in previous works.

%--------------------------------------------------------------------
\paragraph{\textcolor{violet}{Mean Squared Error (MSE)}}
%--------------------------------------------------------------------
One of the commonly used techniques for training a neural network is by minimizing the $L_2$ norm or Mean Squared Error (MSE).
This methods has been widely adopted in various previous studies  \citep{ribli2018improved, lu2022simultaneously, lu2023cosmological}, where the loss function is typically formulated as follows:
\begin{equation}
   \mathcal{L_{\text{MSE}}}= \frac{1}{N_{\theta}}\sum_{i=1}^{N_{\theta}}(t_i-\theta_i)^2.
\end{equation}

Here $N_{\theta}$ represents the number of cosmological parameters, $t$ denotes the summary statistics, and $\theta$ corresponds to the cosmological parameters. 
Minimizing the $L_{2}$-norm is equivalent to training the model to estimate the mean of the posterior distribution. We prove this statement in \autoref{Sec:appendix_Mean Square Error}. However, it is important to note that this approach does not guarantee the recovery of maximally informative summary statistics as it ignores the shape, spread, and correlation structure of the posterior distribution. Indeed, two posteriors can have the same mean while exhibiting different posterior distributions: the posterior mean is not guaranteed to be a sufficient statistic. 

%--------------------------------------------------------------------
\paragraph{\textcolor{violet}{Mean Absolute Error (MAE)}}
%--------------------------------------------------------------------
Another commonly used approach involves minimizing the $L_1$-norm or Mean Absolute Error (MAE). In this approach, the loss function is defined as:
\begin{equation}
    \mathcal{L_{\text{MAE}}}=\frac{1}{N_{\theta}}\sum_{i=1}^{N_{\theta}}|t_i-\theta_i|.
\end{equation}
where $t$ represents the summary statistics and $\theta$ denotes the cosmological parameters.
In \autoref{Sec:appendix_Maximum Absolute Error}, we demonstrate that minimizing this loss function is equivalent to training the model to estimate the median of the posterior distribution. While extensively employed in various previous studies \citep{2018PhRvD..97j3515G, fluri2018cosmological, ribli2019weak}, it is important to note that this loss suffers the same pathology as the MSE loss.
%--------------------------------------------------------------------
\paragraph{\textcolor{violet}{Variational Mutual Information Maximization (VMIM)}}
%--------------------------------------------------------------------
This technique was first introduced for cosmological inference problems by \citet{jeffrey2021likelihood}. This approach aims to maximize the mutual information $I(\bm{t}, \bm {\theta})$ between the cosmological parameters $\bm{\theta}$ and the summary statistics $\bm{t}$. Maximizing this mutual information helps construct sufficient summary statistics, as $\bm{t}$ is sufficient for the parameters $\bm{\theta}$ if and only if $I(\bm{x},\bm{\theta}) =  I(\bm{t},\bm{\theta})$ by definition.

In the VMIM approach, the loss function is defined as:
\begin{equation}\label{Eq:Loss_vmim}
    \mathcal{L_{\text{VMIM}}}=- \log{q(\bm {\theta} |F_{\bm {\varphi}}(\bm {x}) ; \bm{\varphi}')}.
\end{equation}
Here, $q(\bm {\theta} |F_{\bm {\varphi}}(\bm {x});\bm{\varphi'})$ represents a variational conditional distribution, where $\bm{\theta}$ corresponds to the data vector of the cosmological parameters, and $\bm{\varphi'}$ to the parameters characterizing the variational conditional distribution itself. $F_{\bm {\varphi}}$ denotes the compression network of parameters $\bm {\varphi}$, used to extract the summary statistics $\bm{t}$ from the original high-dimensional data vector $\bm{x}$, such that  $\bm{t}=F_{\bm {\varphi}}(\bm{x})$.
In order to understand the significance of this loss function, it is necessary to start by considering the mathematical definition of mutual information $I(\bm{t}, \bm {\theta})$:
\begin{align}\label{Eq:mutual_information}
    I(\bm{t}, \bm {\theta}) &= D_{KL}(p(\bm {t}, \bm {\theta})||p(\bm {t})p(\bm {\theta})) \\ \nonumber
    &= \int d\bm{\theta} d\bm{t} p(\bm t, \bm \theta)\log{\left( \frac{ p(\bm {t}, \bm {\theta})}{ p(\bm {t}) p(\bm {\theta})} \right)} \\ \nonumber
    &= \int d\bm{\theta} d\bm{t} p(\bm t, \bm {\theta})\log{\left( \frac{ p(\bm {\theta} | \bm {t} )}{ p(\bm {\theta})} \right)} \\ \nonumber
        &= \int d\bm{\theta} d\bm{t} p(\bm t, \bm {\theta})\log{p(\bm {\theta} | \bm {t} )} - \int d\bm{\theta}  d\bm{t} p(\bm t, \bm {\theta})\log{p(\bm {\theta})} \\ \nonumber
    &= \int d\bm{\theta} d\bm{t} p(\bm t, \bm {\theta})\log{p(\bm {\theta} | \bm {t} )} - \int d\bm{\theta} p(\bm {\theta})\log{p(\bm {\theta})} \\ \nonumber
    &= \mathbb{E}_{p(\bm {t}, \bm {\theta})} [\log{p(\bm {\theta} | \bm {t} )}]- \mathbb{E}_{p(\bm {\theta})} [\log{p(\bm {\theta})}] \\ \nonumber
    &= \mathbb{E}_{p(\bm {t}, \bm {\theta})} [\log{p(\bm {\theta} | \bm {t} )}]- H(\bm {\theta});
\end{align}
in the above equation, $D_{KL}$ is the Kullback-Leibler divergence \citep{kullback1951information}, $p(\bm {t}, \bm {\theta})$ is the joint probability distribution of summary statistics and cosmological parameters, and $H(\bm {\theta})$ represents the \textit{entropy} of the distribution of cosmological parameters.  
Essentially, mutual information measures the amount of information contained in the summary statistics $\bm t$ about the cosmological parameters $\bm \theta$.
The goal is to find the parameters of the network $\bm {\varphi}$ that maximize the mutual information between the summary and cosmological parameters:
\begin{equation}
   \bm {\varphi}^*= \operatorname*{argmax}_{\bm {\varphi}} I(F_{\bm {\varphi}}(\bm{x}), \bm {\theta}).
\end{equation}
However, the mutual information expressed in \autoref{Eq:mutual_information} is not tractable since it relies on the unknown posterior distribution. To overcome this limitation, various approaches that rely on tractable bounds have been developed, enabling the training of deep neural networks to maximize mutual information. In this study, we adopt the same strategy used by \citet{jeffrey2021likelihood}, which involves using the variational lower bound \citep{barber2003information}:
\begin{equation}\label{Eq:variational_lower_bound}
    I(\bm{t}, \bm{\theta}) \ge \mathbb{E}_{p(\bm {t}, \bm {\theta})} [\log{q(\bm {\theta} |\bm{t} ; \bm{\varphi}')}]- H(\bm {\theta}).
\end{equation}
Here, the variational conditional distribution $\log{q(\bm {\theta} |\bm{t}; \bm{\varphi}')}$ is introduced to approximate the true posterior distribution $p(\bm{\theta}|\bm {t})$. 
As the entropy of the cosmological parameters remains constant with respect to $\bm {\varphi}$, the optimization problem based on the lower bound in \autoref{Eq:variational_lower_bound} can be formulated as:
\begin{equation}
    \operatorname*{argmax}_{\bm {\varphi}, \bm {\varphi}'}\mathbb{E}_{p(\bm {x}, \bm {\theta})} [\log{q(\bm {\theta} |F_{\bm {\varphi}}(\bm {x}) ; \bm{\varphi}')}], 
\end{equation}
yielding \autoref{Eq:Loss_vmim}.

Given the fundamental principles of deep learning, such as having a sufficiently flexible neural network and a sufficiently large dataset, this method is capable of constructing sufficient statistics by design.

One may see a similarity between VMIM and NPE with an information bottleneck (IB) \citep{tishby2000informationbottleneckmethod, alemi2019deepvariationalinformationbottleneck}, as both approaches aim to maximize the mutual information $I(\bm{t}; \bm{\theta})$ between parameters and summary statistics. However, while IB introduces an adaptive trade-off between relevance and compression 
\begin{align}
    \operatorname*{argmax}_t\underbrace{I(\bm{t}; \bm{\theta})}_{\text{relevance}} - \beta \underbrace{I(\bm{t}; \bm{x})}_{\text{compression}},
\end{align}
VMIM achieves a similar effect by directly minimizing $I(\bm{t};\bm{x})$ through dimensional constraints. By enforcing $dim(\bm{t}) \ll dim(\bm{x})$, VMIM ensures that the representation selectively retains information, implicitly reducing irrelevant details without the need to fine-tune the trade-off parameter $\beta$.
%--------------------------------------------------------------------
\paragraph{\textcolor{violet}{Gaussian Negative Log-Likelihood (GNLL)}}
%--------------------------------------------------------------------
Recognizing that the aleatoric uncertainty on different cosmological parameters will vary, a third
class of inverse variance weighted MSE was proposed in \cite{fluri2018cosmological} with the idea of
ensuring that each parameter contributes fairly to the overall loss by taking into account its uncertainty. The loss function typically takes the following form:
\begin{equation}
    \mathcal{L_{\text{GNLL}}} = \frac{1}{2} \log(|\bm{\Sigma}|) + \frac{1}{2}(\bm{t} - \bm{\theta})^{\top} \Sigma^{-1} (\bm{t} - \bm{\theta}) ,
\end{equation}
where $\bm{t}$ is the summary statistics and $\bm{\Sigma}$ is the covariance matrix representing the uncertainty on the cosmological parameters $\bm{\theta}$. Both $\bm{t}$ and $\bm{\Sigma}$ can be outputs of the compression network, i.e. $F_{\bm{\varphi}}(\bm{x})=(\bm{t}, \bm{\Sigma})$.  But only the mean is kept as summary statistics.

One recognizes here the expression of a Gaussian probability function, and this expression can thus be related to the VMIM case by simply assuming a Gaussian distribution as the variational approximation for the posterior $q(\bm{\theta} | \bm{x}) = \mathcal{N}(\bm{\theta}; \bm{t}, \bm{\Sigma})$. 
We demonstrate in \autoref{Sec:appendix_gnll} that under this loss function the summary $\bm{t}$ extracted by the neural network is, similarly to the MSE case, only an estimate of the mean of the posterior distribution, which is not guaranteed to be sufficient. Note that the summary statistics obtained by GNLL should thus be the same as the ones obtained from MSE but at a greater cost since GNLL requires optimizing the mean and the coefficients of the covariance matrix of the Gaussian jointly. We further find in practice that optimizing this loss instead of the MSE is significantly more challenging (as we will illustrate in the results section), which we attribute to the log determinant term of the loss being difficult to optimize.

% }
%--------------------------------------------------------------------
\paragraph{\textcolor{violet}
{Information Maximising Neural Networks (IMNNs)}} 
A different approach has been proposed by \cite{charnock2018automatic} and further explored in \citet{Makinen_2021, Makinen_2022}. They implemented the Information Maximizing Neural Network (IMNN), a neural network trained on forward simulations designed to learn optimal compressed summaries, in circumstances where the likelihood function is intractable or unknown.
Specifically, they propose a new scheme to find optimal non-linear data summaries by using the Fisher information to train a neural network. 
Inspired by the MOPED algorithm \citep{heavens2000massive}, the IMNN is a  transformation $f$ that maps the data to compressed summaries: $f: \textbf{x} \to \textbf{t}$ while conserving the Fisher information.  
The loss function takes the following form:
\begin{equation}\label{eq:loss_imnn}
    \mathcal{L}_{\text{IMNN}} = -  \ln \text{det}(\textbf{F})+ r_{\Sigma},
\end{equation}
where $\textbf{F}$ is the Fisher matrix, and $r_{\Sigma}$ is a regularization term typically dependent on the covariance matrix, introduced to condition the variance of the summaries.
Since computing the Fisher matrix requires a large number of simulations, they proceed as follows: a large number of simulations with the same fiducial cosmology but different initial random conditions are fed forwards through the network. The summaries from these simulations are combined to compute the covariance matrix. Additionally, the summaries from simulations created with different fixed cosmologies are used to calculate the derivative of the mean of the summary with respect to the parameter.\footnote{The method of finite differences is necessary when the framework in which the code is implemented does not support automatic differentiation.} Finally, the covariance and the mean derivatives are combined to obtain the Fisher matrix. We leave the comparison of this scheme on our weak lensing simulations to a future work, due to the memory constraints of available devices and the large batch sizes required for computing \autoref{eq:loss_imnn}.

\citet{fluri2021cosmological, fluri2022full} make use of a re-factored Fisher-based loss in their weak lensing compression. They rearrange \autoref{eq:loss_imnn} such that the output summary does not follow a Gaussian sampling distribution:
\begin{equation}\label{eq:loss_gfim}
\mathcal{L_{\text{Fluri}}}=\log{\text{det}(\bm{\Sigma}_{\bm{\theta}}(\bm{t})})-2\log{\left|\text{det}\left(\frac{\partial \Psi_{\bm{\theta}}(\bm{t})}{\partial \bm{\theta}}\right)\right|},
\end{equation}
where $\bm{\Sigma}$ is the covariance matrix, and $\Psi_{\bm{\theta}}(\bm{t})=\mathbb{E}_{p(\bm {x}|\bm {\theta})}[\bm{t}]$.
This loss function is equivalent to the log-determinant of the inverse Fisher matrix used in \autoref{eq:loss_imnn}. Implementing this loss in \citet{fluri2021cosmological, fluri2022full} did not provide large improvements in cosmological constraints beyond the power spectrum, likely due to the smaller number of training simulations.

\section{The \sbilens\ framework}\label{Sec:the SBILens framework}
To investigate the questions above, we have developed the Python package \href{https://github.com/DifferentiableUniverseInitiative/sbi_lens}{\sbilens}, which provides a weak-lensing differentiable simulator based on a lognormal model. \sbilens\ enables the sampling of convergence maps in a tomographic setting while considering the cross-correlation between different redshift bins.
\subsection{Lognormal Modeling}\label{Sec:Lognormal Modeling}
For various cosmological applications, the non-Gaussian field can be modeled as a lognormal field \citep{coles1991lognormal,bohm2017bayesian}.
This model offers the advantage of generating a convergence field rapidly while allowing the extraction of information beyond the two-point statistics. 
Although studies demonstrated that this model fails in describing the 3D field \citep{klypin2018density}, it properly describes the 2D convergence field \citep{clerkin2017testing, xavier2016improving}.
Assuming a simulated Gaussian convergence map $\kappa_g$, whose statistical properties are fully described by its power spectrum $C_{\ell}$, we know that this model is not a suitable representation of late-time and more evolved structures. One potential solution is to find a transformation $f(\kappa_g)$ of this map that mimics the non-Gaussian features in the convergence field. In doing so, it is crucial to ensure that the transformed map maintains the correct mean and variance, effectively recovering the correct two-point statistics.
Denoting $\mu$ and $\sigma_g^2$ the mean and covariance matrix of $\kappa_g$ respectively, we can define the transformed convergence $\kappa_{ln}$ as a shifted lognormal random field:
\begin{equation}\label{Eq:log_norm_kappa}
    \kappa_{ln}=e^{\kappa_{g}}-\lambda, 
\end{equation}
where $\lambda$ is a free parameter that determines the shift of the lognormal distribution. The convergence $\kappa$ in a given redshift bin is fully determined by the shift parameter $\lambda$, the mean $\mu$ of the associated Gaussian field $\kappa_g$, and its variance $\sigma_{g}^2$.
The correlation of the lognormal field, denoted as $\xi_{ln}$, is also a function of these variables and is related to $\xi^{ij}_g$ through the following equations:
\begin{align}
    \xi^{ij}_{ln}(\theta) & \equiv \lambda_i \lambda_j (e^{ \xi^{ij}_g(\theta)}-1) \nonumber \\ 
    \xi^{ij}_g(\theta)&=\log{\left[ \frac{\xi^{ij}_{ln}(\theta)}{\lambda_i \lambda_j}+1\right ]}. \label{Eq:log_norm_corr}
\end{align}
Here $i$ and $j$ define a pair of redshift bins.
The parameter $\lambda$, also known as \textit{minimum convergence
parameter}, defines the lowest values for all possible values of $\kappa$.
The modeling of the shift parameter can be approached in various ways. For example, it can be determined by matching moments of the distribution \citep{xavier2016improving} or by treating it as a free parameter \citep{hilbert2011cosmic}. In general, the value of $\lambda$ depends on the redshift, cosmology, and the scale of the field at which smoothing is applied.

While it is straightforward to simulate a single map, if we want to constrain the convergence map in different redshift bins, an additional condition must be met. The covariance of the map should recover the correct angular power spectrum:
\begin{equation}\label{power_spectrum_definition}
    \left \langle \tilde{\kappa}^{(i)}_{ln} (\ell)\tilde{\kappa}^{*(j)}_{ln}(\ell')\right \rangle =C^{ij}_{ln}(\ell)\delta^{K}(\ell-\ell'),
\end{equation}
where $ C^{ij}_{ln}(\ell)$ is the power spectrum of $\kappa_{ln}$ in Fourier space, defined as:
\begin{equation}\label{Eq:log_norm_cls}
    C^{ij}_{ln}(\ell)=2\pi \int_0^{\pi} d\theta \sin{\theta}P_{\ell}(\cos{\theta})\xi^{ij}_{ln}(\theta)
\end{equation}
and $P_{\ell}$ is the Legendre polynomial of order $\ell$. 
Using the lognormal model, we can simultaneously constrain the convergence field in different redshift bins while considering the correlation between the bins, as described by \autoref{Eq:log_norm_corr}.

 In the \sbilens\ framework, the sampling of the convergence maps can be described as follows. 
First, we define the survey in terms of galaxy number density, redshifts, and shape noise.  Then, we compute the theoretical auto-angular power spectrum $C^{ii}(\ell)$ and cross-angular power spectrum $C^{ij}(\ell)$ for each tomographic bin. These theoretical predictions are calculated using the public library \href{https://github.com/DifferentiableUniverseInitiative/jax_cosmo}{\texttt{jax-cosmo}} \citep{Campagne_2023}. 
Next, we project the one-dimensional $C(\ell)$ onto two-dimensional grids with the desired final convergence map size. Afterwards, we compute the Gaussian correlation functions $\xi^{ij}_g(\theta)$ using \autoref{Eq:log_norm_corr}.
 To sample the convergence field in a specific redshift bin while considering the correlation with other bins, we use \autoref{power_spectrum_definition}. 
We construct the covariance matrix $\bm{\Sigma}$ of the random field $\bm{\kappa}$, where $\bm{\kappa}$ represents the vector of convergence maps at different redshifts as follows:
 \begin{equation}
    \bm{\Sigma}= 
    \begin{pmatrix}
    C_{\ell}^{11} & C_{\ell}^{12} & \cdots & C_{\ell}^{1n} \\
    C_{\ell}^{21} & C_{\ell}^{22} & \cdots & C_{\ell}^{2n} \\
    \vdots  & \vdots  & \ddots & \vdots  \\
    C_{\ell}^{n1} & C_{\ell}^{n2} & \cdots & C_{\ell}^{nn} 
    \end{pmatrix}.
\end{equation}
To sample more efficiently, we perform an eigenvalue decomposition of $\bm{\Sigma}$ to obtain a new matrix $\tilde{\bm{\Sigma}}$:
\begin{equation}
    \tilde{\bm{\Sigma} }=\bm{Q}\bm{\Lambda}^{1/2}\bm{Q}^{T}
\end{equation}
where $\bm{Q}$ and $\bm{\Lambda}$ are the eigenvectors and eigenvalues of $\bm{\Sigma}$, respectively.
Next, we sample the Gaussian random maps $\bm{\kappa_g}$ using the equation:
\begin{equation}
     \bm{\kappa_g}=\hat{\bm{Z}}\cdot\tilde{\bm{\Sigma} }
\end{equation}
where $\hat{\bm{Z}}$ represents the Fourier transform of the latent variables of the simulator.
Finally, we transform the Gaussian map $\kappa_g$ into a lognormal field using \autoref{Eq:log_norm_kappa}.
 \\
To ensure that we recover the correct auto- and cross-power spectra, we compare the results from our simulations to theoretical predictions for different tomographic bin combinations. We show the results in \autoref{fig:psconvergence_maps}. 
%------------------------------------------------------------------
%------------------------------------------------------------------
%--------------------------------------------------------------------
\subsection{Data generation}
\begin{figure*}
    \begin{center}
    \includegraphics[width=\textwidth]{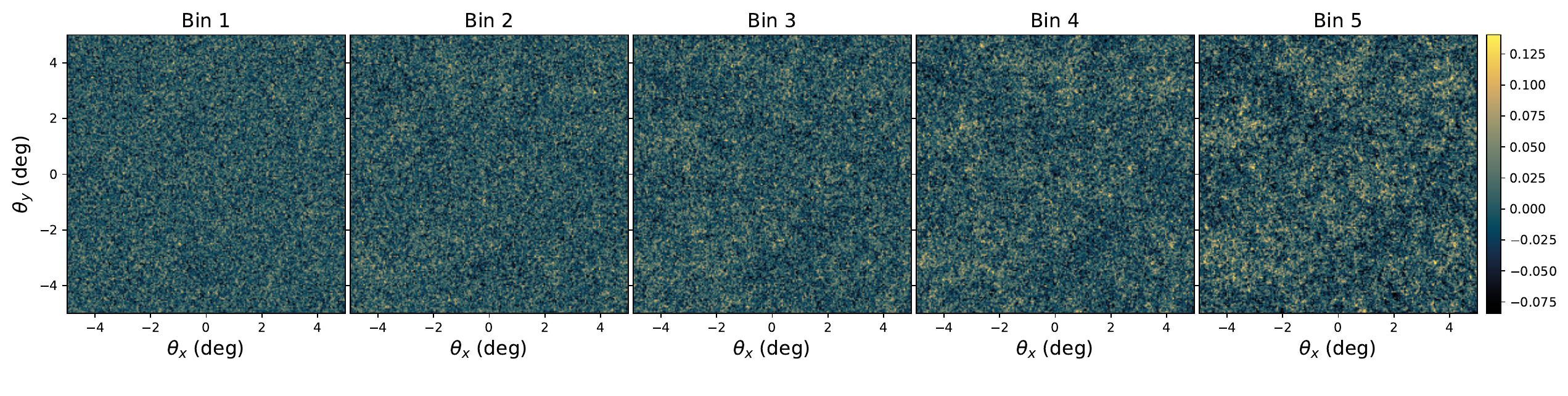}
    \caption{
     Example of convergence maps simulated using the \sbilens\ 
 package.
    }
     \label{fig:convergence_maps}
     \end{center}
\end{figure*}

%--------------------------------------------------------------------
%--------------------------------------------------------------------
%--------------------------------------------------------------------
Our analysis is based on a standard flat $w$CDM cosmological model,  which includes the following parameters: the baryonic density fraction $\Omega_b$, the cold dark matter density fraction $\Omega_c$, the Hubble parameter $h_0$, the spectral index $n_s$, the amplitude of the primordial power spectrum $\sigma_8$ and the dark energy parameter $w_0$. The priors used in the simulations and in the inference process are listed in \autoref{tab:prior}, following \citet{zhang2022transitioning}.
To simulate our data, we develop the \sbilens\ package, which employs a lognormal model to represent the convergence maps, as explained in the previous section. Specifically, the package uses the public library \href{https://github.com/DifferentiableUniverseInitiative/jax_cosmo}{\texttt{jax-cosmo}} to compute the theoretical power- and cross-spectra. The computation of the lognormal shift parameter is performed using the \texttt{Cosmomentum} code \citep{friedrich2018density, friedrich2020primordial}, which utilizes perturbation theory to compute the cosmology-dependent shift parameters. In \texttt{Cosmomentum} the calculation of the shift parameters assumes a cylindrical window function, while our pixels are rectangular. Following \citet{boruah2022map}, we compute the shift parameters at a characteristic scale, $R=\Delta L/\pi$, where $\Delta L$ represents the pixel resolution. \\
For each redshift bin, we tested the dependency of the shift parameter $\lambda$ on various cosmological parameters. Specifically, we investigated how the value of $\lambda$ changed when varying a specific cosmological parameter while keeping the others fixed.
Our findings revealed that the parameters $\Omega_b$, $h_0$, and $n_s$ had almost no significant impact on $\lambda$. As a result, we computed the shift parameters for each redshift using the fiducial cosmology values of $\Omega_b$, $h_0$, and $n_s$.
To account for the cosmology dependence of $\lambda$ on $\Omega_c$, $\sigma_8$, and $w_0$, we calculated the shift for various points in the cosmological parameter space and then interpolated the shift values for other points in the parameter space.
Each map is reproduced on a regular grid with dimensions of $256 \times 256$ pixels and covers an area of $10\times 10$ deg$^2$.  An example of a tomographic convergence map simulated using the \sbilens\ package is shown in \autoref{fig:convergence_maps}. 
%--------------------------------------------------------------------
% ############# PRIOR COSMO TABLE  #############
%--------------------------------------------------------------------
\begin{table}
	\begin{center}
    	\begin{tabular}{lcc} 
    		\hline \hline
    		Parameter  & Prior & Fiducial value \\
    		$\Omega_c$ & $\mathcal{N}_T$ (0.2664, 0.2) & 0.2664 \\
    		$\Omega_b$ & $\mathcal{N}$ (0.0492, 0.006) & 0.0492 \\
    		$\sigma_8$ & $\mathcal{N}$ (0.831, 0.14) & 0.831 \\
    		$h_0$ & $\mathcal{N}$ (0.6727, 0.063) & 0.6727\\
    		$n_s$ & $\mathcal{N}$ (0.9645, 0.08) & 0.9645 \\
    		$w_{0}$ &  $\mathcal{N}_T$ (-1.0, 0.9) &  -1.0 \\
    		\hline
    	\end{tabular}
        \caption{ Prior and fiducial values used for the analyses. 
        The symbol $\mathcal{N}_T$ represents a Truncated Normal distribution. The lower bound of the support for the $\Omega_c$ distribution is set to zero, while the lower and upper bounds for the $w_0$ distribution are set to -2.0 and -0.3, respectively.}
	    \label{tab:prior}
    \end{center}
\end{table}
%------------------------------------------------------------------
%------------------------------------------------------------------
%------------------------------------------------------------------
\subsection{Noise and survey setting}
\begin{figure}
    \centering
    \includegraphics[width=\columnwidth]{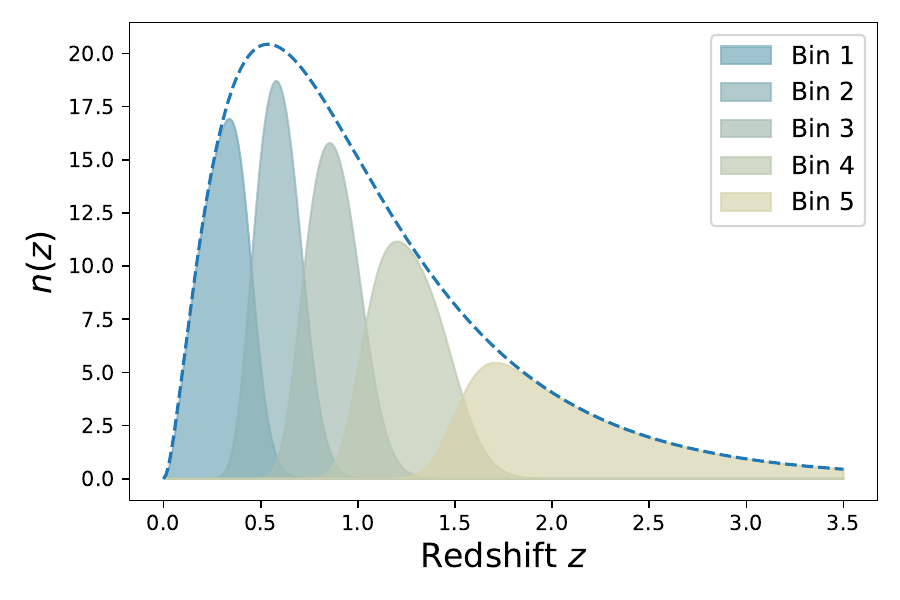}
    \caption{
     Source sample redshift distributions for each tomographic bin for LSST Y10. The number density on the y-axis is shown in arcmin$^{-2}$.
    }
     \label{fig:redshift_distribution}
\end{figure}

%--------------------------------------------------------------------
%--------------------------------------------------------------------
%--------------------------------------------------------------------
We conduct a tomographic study to reproduce the redshift distribution and the expected noise for the LSST Y10 data release.
Following \citet{zhang2022transitioning}, we model the underlying redshift distribution using the parametrized Smail distribution \citep{smail1995deep}:
\begin{equation}
    n(z) \propto z^2 \exp{-(z/z_0)^{\alpha}},
\end{equation}
with $z_0=0.11$ and $\alpha=0.68$. We also assume a photometric redshift error $\sigma_z=0.05(1+z)$ as defined in the LSST DESC Science Requirements Document (SRD, \citet{mandelbaum2018lsst}).
The galaxy sources are divided into five tomographic bins, each containing an equal number of sources $N_s$, computed from the average galaxy number density $n_{gal}$ and the survey pixel area $A_{pix}$. 
For each redshift bin, we assume Gaussian noise with mean zero and variance given by
 \begin{equation}
     \sigma^2_n= \frac{\sigma_e^2}{N_s}.
 \end{equation} 
\autoref{fig:redshift_distribution} illustrates the resulting source redshift distribution, and \autoref{tab:survey_spec} provides a summary of the survey specifications.
%--------------------------------------------------------------------
% ############# SURVEY SPECIFICATION TABLE  #############
%--------------------------------------------------------------------
\begin{table}
	\begin{center}
    	\begin{tabular}{lc} 
            \hline \hline
    		Redshift binning & 5 bins \\
    		Redshift distribution ($z_{0}, \alpha$) & (0.11, 0.68)  \\
    		Number density $n_{gal}$ & 27 arcmin$^{-2}$ \\
    		Shape noise $\sigma_e$ & 0.26 \\
    		Redshift error $\sigma_z$ &0.05(1+z)  \\
                Pixel area $A_{pix}$ &5.49 arcmin$^2$ \\
    		\hline
    	\end{tabular}
     	\caption{ LSST Y10 source galaxy specifications in our analysis. All values are based on the LSST DESC Science Requirements Document.}
	    \label{tab:survey_spec}
    \end{center}
\end{table}
%--------------------------------------------------------------------
% ############# PLOT REDSHIFT DISTRIBUTION  #############
%--------------------------------------------------------------------
%--------------------------------------------------------------------
% ############# PLOT CONVERGENCE MAPS  #############
%--------------------------------------------------------------------
%--------------------------------------------------------------------
%--------------------------------------------------------------------
%--------------------------------------------------------------------
%------------------------------------------------------------------
\section{Experiment}\label{Sec:experiment}
In the following section, we illustrate the inference strategy for the two approaches under investigation: the Bayesian forward modeling and the map-based inference based on implicit inference.
Additionally, we conduct a power spectrum study. Indeed, as discussed in \autoref{Sec:Lognormal Modeling}, lognormal fields offer the advantage of rapidly generating convergent fields while accounting for non-Gaussianities. To emphasize this claim further, along with the full-field analysis, we include a power spectrum analysis. This analysis demonstrates that there is indeed a gain of information when adopting a full-field approach. 
%--------------------------------------------------------------------
%--------------------------------------------------------------------
%--------------------------------------------------------------------
\subsection{Explicit Inference}
%--------------------------------------------------------------------
%--------------------------------------------------------------------
%--------------------------------------------------------------------
\subsubsection{Full-field inference with BHMs}
%--------------------------------------------------------------------
%--------------------------------------------------------------------
%--------------------------------------------------------------------
The explicit joint likelihood $p(\bm{x}|\bm{z},\bm{\theta})$ provided by an explicit forward model is the key ingredient for conducting explicit full-field inference. In the following, we describe how we build this likelihood function based on the forward model described in \autoref{Sec:Lognormal Modeling}.

The measurement of convergence for each pixel and bin will differ from real observations due to noise. This is taken into consideration in the likelihood. Specifically, for LSST Y10, the number of galaxies for each pixel should be sufficiently high so that, according to the central limit theorem, we can assume the observation is characterized by Gaussian noise, with $\sigma_n^2=\sigma_e^2/N_s$, where $N_s$ represents the total number of source galaxies per bin and pixel. Given $\sigma_n^2$ the variance of this Gaussian likelihood, its negative log-form can be expressed as:
\begin{equation}
    \mathcal{L}(\bm{\theta}, \bm{z})=
    - \sum_i^{N_{pix}} \sum_{j}^{N_{bins}} \log{
    p(\kappa^{obs}_{i,j}|\kappa_{i,j},\bm{\theta})}
    \propto \sum_i^{N_{pix}} \sum_{j}^{N_{bins}}\frac{[\kappa_{i,j}-\kappa^{obs}_{i,j}]^2}{2\sigma_n^2},
\end{equation}
where $\kappa^{obs}$ refers to the observed noisy convergence maps. This map is fixed for the entire benchmark. 

Since the explicit full-field approach involves sampling the entire forward model, it typically leads to a high-dimensional problem, requiring more sophisticated statistical techniques. To sample the posterior distribution $p(\bm{\theta},\bm{z}|\bm{x})$, we use a Hamiltonian Monte Carlo (HMC) algorithm. 
Specifically, we employ the \texttt{NUTS} algorithm \citep{hoffman2014no}, an adaptive variant of HMC implemented in \href{https://github.com/pyro-ppl/numpyro}{\texttt{NumPyro}} \citep{phan2019composable, bingham2019pyro}. \\
The HMC algorithm is particularly helpful in high-dimensional spaces where a large number of steps are required to effectively explore the space. It improves the sampling process by leveraging the information contained in the gradients to guide the sampling process. As the code is implemented within the JAX framework, the gradients of the computation are accessible via automatic differentiation. 
%--------------------------------------------------------------------
%--------------------------------------------------------------------
%--------------------------------------------------------------------
\subsubsection{Power spectrum}
%--------------------------------------------------------------------
%--------------------------------------------------------------------
%--------------------------------------------------------------------
To obtain the posterior distribution of the cosmological parameters given the angular power spectra $C_{\ell}$, we assume a Gaussian likelihood with a cosmological-independent covariance matrix:
\begin{equation}
    \mathcal{L}(\bm{\theta})=-\frac{1}{2}[\bm{d}-\bm{\mu}(\bm{\theta})]^{T}\bm{C}^{-1}[\bm{d}-\bm{\mu}(\bm{\theta})].
\end{equation}
To compute the expected theoretical predictions $\bm{\mu}(\bm{\theta})$ we use \href{https://github.com/DifferentiableUniverseInitiative/jax_cosmo}{\texttt{jax-cosmo}}. 
 The covariance matrix $\bm{C}$ of the observables is computed at the fiducial cosmology, presented in \autoref{tab:prior}, using the same theoretical library. Specifically, in {\texttt{jax-cosmo}}, the Gaussian covariance matrix is defined as:
\begin{equation}
    \text{Cov}(C_{\ell},C_{\ell'})=\frac{1}{f_{sky}(2 \ell+1)}\left(C_{\ell}+\frac{\sigma_{\epsilon}^2}{2n_s}\right)\delta^{K}(\ell-\ell'),
\end{equation}
where $f_{sky}$ is the fraction of sky observed by the survey, and $n_s$ is the number density of galaxies. 
To obtain the data vector $\bm{d}$, containing the auto- and the cross-power spectra for each tomographic bin, we use the \href{https://lenstools.readthedocs.io/en/latest/lenstool} {\texttt{LensTools}} package \citep{2016A&C....17...73P} on the fixed observed noisy map. 
To constrain the cosmological parameters, we sample from the posterior distribution using the \texttt{NUTS} algorithm from  \texttt{NumPyro}.
%--------------------------------------------------------------------
%--------------------------------------------------------------------
%--------------------------------------------------------------------
\subsection{Implicit Inference}
%--------------------------------------------------------------------
%--------------------------------------------------------------------
%--------------------------------------------------------------------
%--------------------------------------------------------------------
\subsubsection{Compression strategy}
%--------------------------------------------------------------------
We propose here to benchmark four of the most common loss functions introduced in \autoref{sec:compression_review}, i.e. MSE, MAE, GNLL, and VMIM. For all of them, we use the same convolutional neural network architecture for our compressor: a ResNet-18 \citep{he2016deep}. The ResNet-18 is implemented using Haiku \citep{haiku2020github}, a Python deep learning library built on top of JAX.

While the different compression strategies share the same architecture, the training strategy for VMIM involves an additional neural network. To train the neural compressor under VMIM, we jointly optimize the weights $\bm{\varphi}$ of the neural network $F_{\bm{\varphi}}$ (the compressor) and the parameters $\bm{\varphi}'$ of the variational distribution $q_{\bm{\varphi}'}$. For VMIM, the variational distribution $q(\bm{\theta}|\bm{t};\bm{\varphi}')$ is modeled using a Normalizing Flow (NF). 
After training, we export the results of the neural compressor $F_{\bm{\varphi}}$ but discard the results from the density estimator. Then, we train a new NF to approximate the posterior distribution.
Indeed, as mentioned before, we perform the implicit inference as a two-step procedure. This choice is motivated by the fact that it is difficult to train a precise conditional density estimator when the compressor can still change from iteration to iteration. Hence, the choice to split the problem into two steps: first, the dimensionality reduction, where the density estimation part does not need to be perfect; second, the density estimation itself, which needs to be done very carefully now, but it is much easier because we are in low dimension.

%--------------------------------------------------------------------
%--------------------------------------------------------------------
%--------------------------------------------------------------------
%--------------------------------------------------------------------
\subsubsection{Inference strategy}\label{Sec:Inference_strategy}
%--------------------------------------------------------------------
%--------------------------------------------------------------------
\begin{figure*}[t!]
\centering
\begin{minipage}{8.7cm}
  \centering
  \includegraphics[width=1\linewidth]{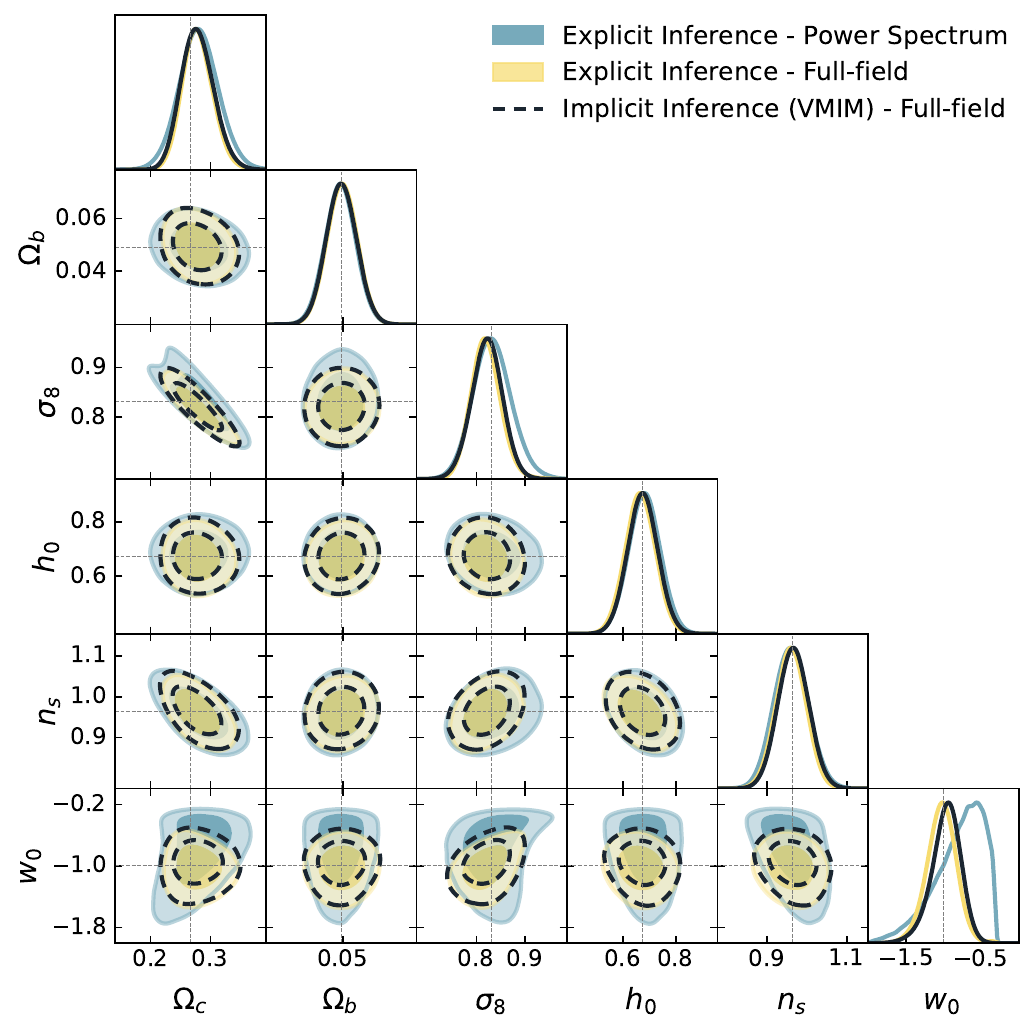}
  \caption{Constraints on the $w$CDM parameter space as found in the LSST Y10 survey setup. The constraints are obtained by applying the $C_{\ell}$ (blue contours), the full-field explicit inference (yellow contours), and the full-field implicit inference strategy using the VMIM compression (black dashed contours), described in \autoref{Sec:experiment}.
    The contours show the $68\%$ and the $95\%$  confidence regions. The dashed lines define the true parameter values.}
  \label{fig:contours_posterior_imp_ex_ps}
\end{minipage}%
\begin{minipage}{0.5cm}
\:
\end{minipage}%
\begin{minipage}{8.7cm}
  \centering
  \includegraphics[width=1\linewidth]{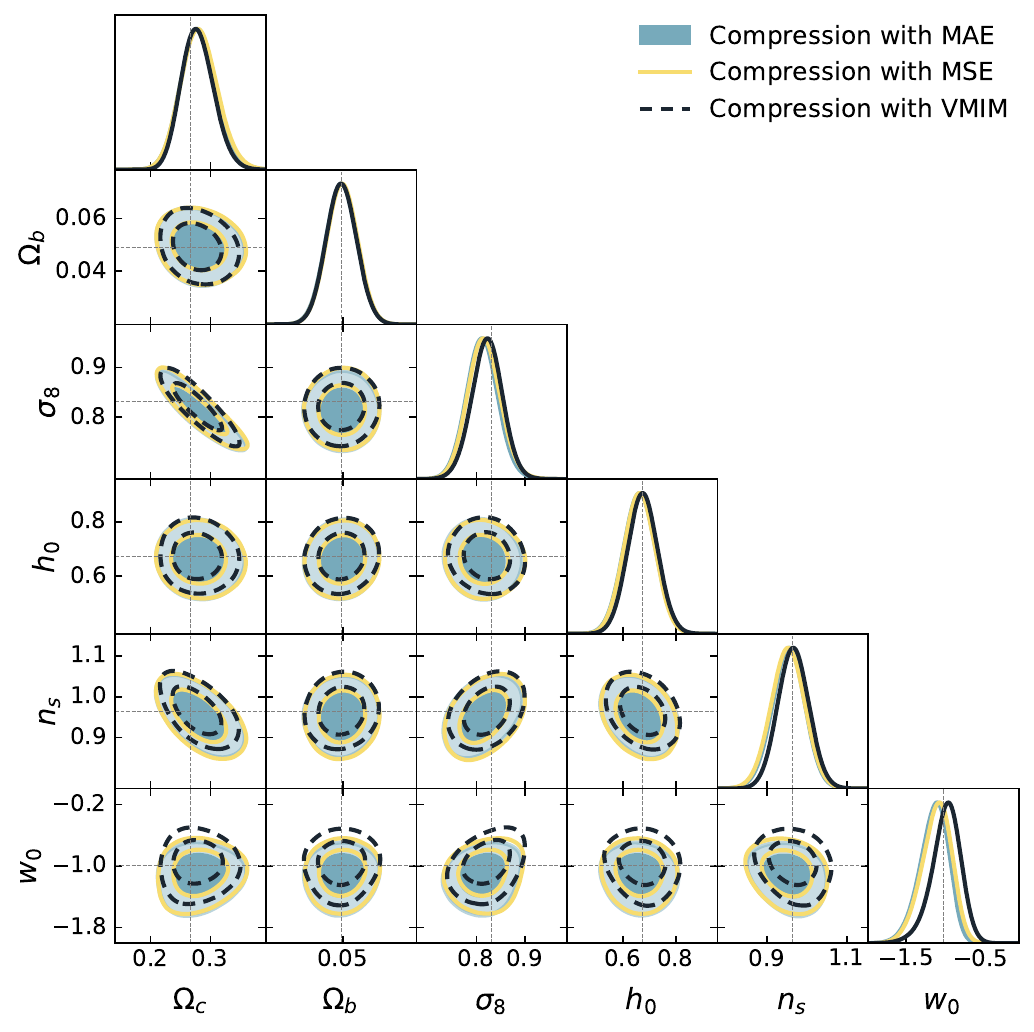}
  \caption{Constraints on the $w$CDM parameter space as found in the LSST Y10 survey setup. The constraints are obtained from three CNN map compressed statistics: the MSE (yellow contours), the MAE (blue contours), VMIM (black dashed contours), described in \autoref{Sec:experiment}. The same implicit inference procedure is used to get the approximated posterior from these four different compressed data. 
The contours show the $68\%$ and the $95\%$  confidence regions. The dashed lines define the true parameter values.}
    \label{fig:contours_posterior_diff_loss}
\end{minipage}
\end{figure*}

\begin{figure}[h]
    \centering
\includegraphics[width=1\linewidth]{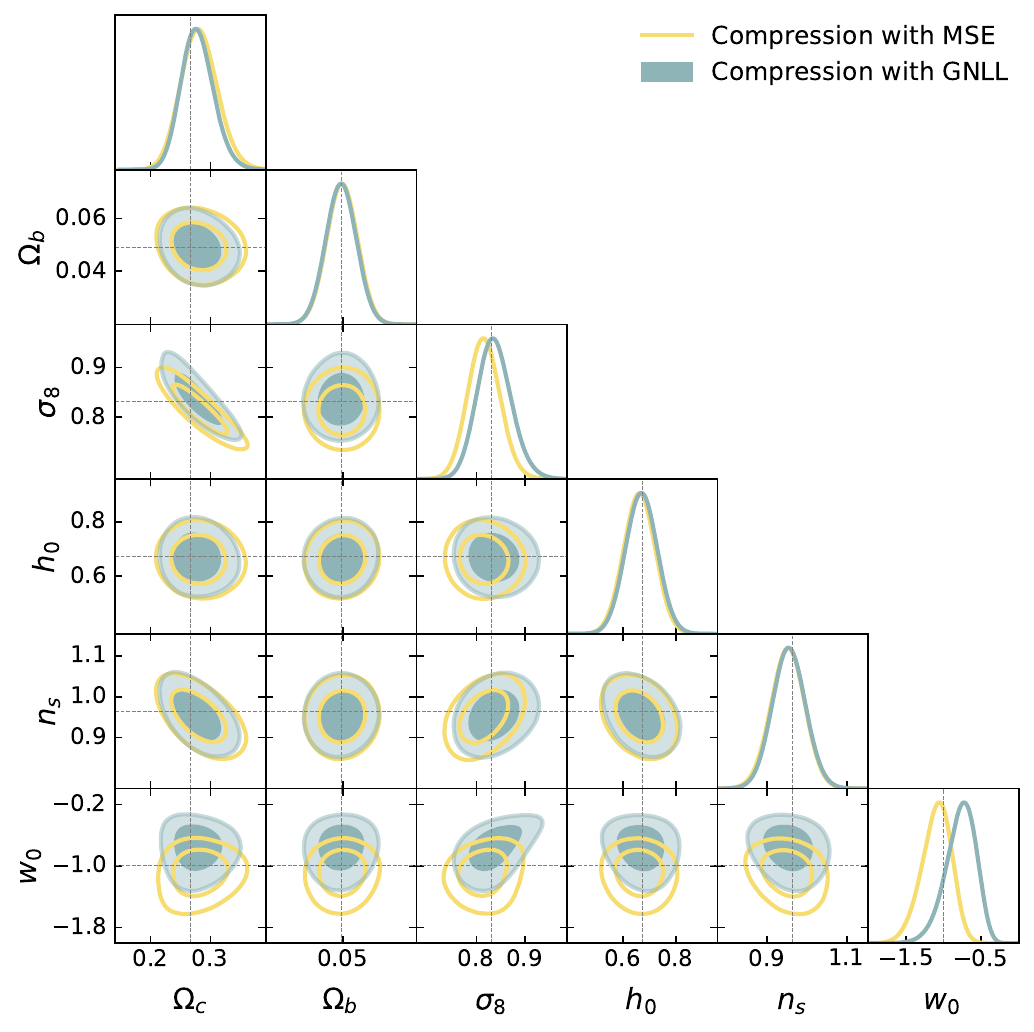}
    \caption{Constraints on the $w$CDM parameter space as found in the LSST Y10 survey setup. We compare the constraints obtained with the GNLL compression (green contours) and MSE compression (yellow contours). These two compressions are supposed to yield the same constraints. The difference is due to optimization issues while training using the GNLL loss function. The same implicit inference procedure is used to get the approximated posterior from these two different compressed data. 
    The contours show the $68\%$ and the $95\%$  confidence regions. The dashed lines define the true parameter values.}
    \label{fig:contour_plot_gnll}
\end{figure}

Based on previous compression we now aim to perform inference. We use implicit inference methods to bypass the problem of assuming a specific likelihood function for the summary statistics $\bm t$.
As mentioned before, implicit inference approaches allow to perform rigorous Bayesian inference when the likelihood is intractable by using only simulations from a black box simulator. \\
We focus on neural density estimation methods where the idea is to introduce a parametric distribution model $q_{\bm{\varphi}'}$ and learn the parameters $\bm{\varphi}'$ from the dataset $(\bm{\theta}_i, \bm{x}_i)_{i=1..N}$ so that it approximates the target distribution. In particular, we focus on NPE, i.e., we aim to directly approximate the posterior distribution.

We use a conditional NF to model the parametric conditional distribution $q_{\bm{\varphi}'} (\bm{\theta} | \bm{t})$, and we optimize the parameters ${\bm{\varphi}'}$ according to the following negative log-likelihood loss function:
\begin{equation}\label{eq:nll}
    \mathcal{L_{\text{NLL}}}=- \log{q_{\bm{\varphi}'}(\bm{\theta}|\bm{t})}.
\end{equation}
In the limit of a large number of samples and sufficient flexibility, we obtain:
\begin{equation}
    q_{\bm{\varphi}'^\ast}(\bm{\theta} | \bm{t}) \approx p(\bm{\theta} | \bm{t}),
\end{equation}
where we indicate with $\bm{\varphi}'^\ast$ the values of $\bm{\varphi}'$ minimizing \autoref{eq:nll}. \\
Finally, the target posterior $p(\bm{\theta} | \bm{t} = \bm{t_0})$ is approximated by $q_{\bm{\varphi}'^\ast}(\bm{\theta} | \bm{t} = \bm{t_0})$, where $\bm{t_0} = F_{\bm{\varphi}}(\bm{x_0})$, i.e., the compressed statistics from the fixed fiducial convergence map $\bm{x_0}$ for a given compression strategy.

For each of the summary statistics, we approximate the posterior distribution using the same NF architecture, namely a RealNVP \citep{realnvp} with 4 coupling layers. The shift and the scale parameters are learned using a neural network with 2 layers of 128 neurons with Sigmoid-weighted Linear Units (SiLU) activation functions \citep{silu}. 
%--------------------------------------------------------------------
%--------------------------------------------------------------------
%--------------------------------------------------------------------
\section{Results}\label{Sec:results}
%--------------------------------------------------------------------
%
\begin{table*}
    \begin{center}
        \begin{tabular}{lcccccc} 
            \hline
            FoM & $C_{\ell}$ & Full-Field (HMC)&  VMIM & MSE & MAE & GNLL   \\
            \hline\hline
            $\Omega_c-  \sigma_8$ & 1222 & 2520 & 2526 & 2043 & 2316 & 1882 \\
            $\Omega_c-  w_0$      & 100  & 198  & 190  & 152  & 162  & 178  \\
            $\sigma_8-  w_0$      & 77   & 176  & 171  & 139  & 149  & 153  \\
            \hline
        \end{tabular}
        \caption{ Figure of Merit (FoM) for different inference strategies: the convergence power spectrum $C_{\ell}$, the explicit full-field through HMC, the CNN map compressed statistics with the MSE, the VMIM, the MAE and the GNLL loss functions. The values of the FoM are inversely proportional to the area of the contours; the larger the FoM, the higher the constraining power.}
        \label{tab:f_o_m}
    \end{center}
\end{table*}
We present the results of the constraints on the full $w$CDM parameter space expected for a survey like LSST Y10. The results are obtained using the simulation procedure outlined in \autoref{Sec:the SBILens framework} and the parameter inference strategy described in \autoref{Sec:Inference_strategy}. The same fiducial map is used for all inference methods.
We begin by presenting the estimators we use to quantify our results and to compare the different approaches. Then we compare the outcomes of three inference procedures: the 2-point statistics, the explicit full-field statistics, and the implicit full-field statistics using CNN summaries. Subsequently, we analyze the impact of the different compression strategies on the final cosmological constraints.
%--------------------------------------------------------------------
\subsection{Result estimators}
We quantify the results by computing the Figure of Merit defined as follows:
\begin{equation}\label{Eq:Figure_of_merit}
    \text{FoM}_{\alpha \beta}=\sqrt{\text{det}(\Tilde{F}_{\alpha \beta})}.
\end{equation}
Here, $\alpha$ and $\beta$ represent a pair of cosmological parameters, and $\Tilde{F}_{\alpha \beta}$ refers to the marginalized Fisher matrix. We calculate $\Tilde{F}_{\alpha \beta}$ as the inverse of the parameter space covariance matrix $C_{\alpha \beta}$, which is estimated from the NF for the implicit inference or the HMC samples for the explicit inference. 
Under the assumption of a Gaussian covariance, the FoM defined in \autoref{Eq:Figure_of_merit} is proportional to the inverse of the $2-\sigma$ contours in the 2-dimensional marginalized parameter space of the $\alpha$ and $\beta$ pair.\\

In addition, from the definition of a sufficient statistics, by comparing the posterior contours to the ground truth (the explicit full-field approach) we can conclude whether a compression strategy is sufficient or not.
%--------------------------------------------------------------------
\subsection{Power spectrum and full-field statistics}
%--------------------------------------------------------------------
We now compare the constraining power of the three approaches described in \autoref{Sec:experiment}: the standard two-point statistics and two map-based approaches, the explicit inference and the implicit inference strategy. 
As outlined before, our interest is to prove that the two map-based approaches lead to very comparable posterior distributions. We present the $68.3\%$ and $95.5\%$ confident regions from the fixed fiducial map for the full $w$CDM parameters in \autoref{fig:contours_posterior_imp_ex_ps}. The contours obtained by the angular $C_{\ell}$ analysis are plotted in blue, the ones for the explicit full-field inference (HMC) in yellow, and those for the implicit full-field inference (VMIM compression and NPE) in black. 
The results are presented in \autoref{tab:f_o_m}.
The remarkably strong agreement between the two posteriors confirms that the two map-based cosmological inference methods yield the same results. The ratio of their FoM, corresponding to $1.00, 1.04, 1.03$, in the $(\Omega_c-\sigma_8); (\Omega_c-w_0); (\sigma_8-w_0$) planes, establishes the validity of the full-field implicit inference strategy. \\
We note that $h$ and $\Omega_b$ are prior dominated and hence are constrained by none of the three approaches. 
Additionally, for the full-field strategies we find that the size of the contours is significantly smaller than the size of the prior distributions adopted.
Moreover, we find that the two-point statistic is sub-optimal in constraining $\Omega_c$, $\sigma_8$, and $w_0$, while the map-based approaches yield much tighter constraints on these parameters. 
We find that the map-based explicit and implicit strategies lead to an improvement in the FoM of $2.06\times$, $1.98\times$, $2.28\times$ and $2.06\times$, $1.90\times$, $2.22\times$, respectively, in the $(\Omega_c-\sigma_8); (\Omega_c-w_0); (\sigma_8-w_0$) plane.
%--------------------------------------------------------------------
\subsection{Optimal compression strategy}
\autoref{fig:contours_posterior_diff_loss} shows our $68.3\%$ and $95.5\%$ constraints from the fixed fiducial map using different compressed summaries.
We compare the constraints obtained from MSE (yellow contours), MAE (blue contours), and VMIM (black dashed contours).
We note that the results obtained using different summaries are generally in agreement with each other, and there are no tensions present for any of the cosmological parameters. 
In \autoref{fig:contour_plot_gnll} we show the difference we obtain between MSE and GNLL compression, which as explained in \autoref{sec:compression_review} should yield the same summary statistics and thus the same constraints. This difference is due to optimization issues. As explained before, the GNLL loss function aims to optimize both the mean and the covariance matrix, which makes the training very unstable compared to the training of MSE.
After investigating different training procedures we conclude that the GNLL loss, because of the covariance matrix, is hard to train and recommend to use of MSE loss if only the mean of the Gaussian is of interest.

We report the marginalized summary constraints in \autoref{tab:summaries}. The results concern the cosmological parameters that are better constrained from weak-lensing: $\Omega_c$, $\sigma_8$, $w_0$.
We note that the VMIM compressed summary statistics prefer values of $\Omega_c$, $\sigma_8$, and $w_0$ that are closer to our fiducial cosmology than those inferred by MSE and MAE. \\
To further quantify these outcomes, we consider the FoM described in \autoref{Eq:Figure_of_merit}. 
The results are presented in \autoref{tab:f_o_m}.
We can see that VMIM yields more precise measurements than MSE, and MAE for all considered parameters.
In particular, the FoM of ($\Omega_c$, $\sigma_8$) is improved by $1.24\times$, and $1.1\times$ compared to MSE, and MAE respectively;  the FoM of ($\Omega_c$, $w_0$) is improved by $1.25\times$, and $1.17\times$ from the MSE, and MAE; the FoM of ($\sigma_8$, $w_0$) is improved by $1.23\times$, and $1.15\times$ from the MSE, and MAE. 

%--------------------------------------------------------------------
% ############# FIGURE OF MERIT TABLE  #############
%--------------------------------------------------------------------
%--------------------------------------------------------------------
%  ############# SUMMARY TABLE  #############
%--------------------------------------------------------------------
\begin{table*}
    \begin{center}
        \begin{tabular}{|l|l|l|l|l|l|}
            \hline
             & Full-Field (HMC) & VMIM & MSE & MAE &GNLL \\
            \hline 
            \hline
            $\Omega_c$ & $0.275^{+0.026}_{-0.024}$ & $0.274^{+0.026}_{-0.025}$  & $0.283^{+0.031}_{-0.029}$  & $0.279^{+0.030}_{-0.028}$ & $0.275^{+0.029}_{-0.025}$\\
            \hline
            $\Omega_b$ & $\left( 49.6^{+5.9}_{-6.0} \right) \times 10^{-3}$ & $ \left( 49.9^{+6.0}_{-6.4} \right) \times 10^{-3}$ &  $\left( 49.2\pm 6.1 \right) \times 10^{-3}$ &  $\left( 50.1^{+5.9}_{-6.1} \right) \times 10^{-3}$ & $\left( 49.5^{+5.7}_{-6.2} \right) \times 10^{-3}$ \\
            \hline
            $\sigma_8$ & $0.816^{+0.032}_{-0.029}$ & $0.819^{+0.031}_{-0.029}$  & $0.808^{+0.034}_{-0.032}$  &  $0.808^{+0.032}_{-0.030}$ &  $0.834^{+0.036}_{-0.035}$ \\
            \hline
            $w_0$      & $-1.01^{+0.19}_{-0.20}$ & $-1.00^{+0.20}_{-0.22}$  & $-1.13^{+0.23}_{-0.22}$ &   $-1.16 \pm 0.22$  &  $-0.73^{+0.19}_{-0.22}$ \\
            \hline
            $h_0$      & $0.664^{+0.059}_{-0.056}$& $0.666^{+0.061}_{-0.057}$ & $0.662^{+0.056}_{-0.057}$  & $0.664^{+0.056}_{-0.058}$  & $0.670^{+0.059}_{-0.061}$\\
            \hline
            $n_s$      & $0.960\pm 0.038$ & $0.963^{+0.041}_{-0.038}$ &  $0.956 \pm 0.041$ &  $0.955^{+0.039}_{-0.041}$  & $0.954\pm 0.042$\\
            \hline
        \end{tabular}
        \caption{Summary of the marginalized parameter distributions.}
        \label{tab:summaries}
  \end{center}
\end{table*}
%--------------------------------------------------------------------
%--------------------------------------------------------------------
\section{Conclusion}\label{Sec:conclusion}

Implicit inference typically involves two steps. The first step is the automatic learning of an optimal low-dimensional summary statistic, and the second step is the use of a neural density estimator in low dimensions to approximate the posterior distribution. In this work, we evaluated the impact of different neural compression strategies used in the literature on the final posterior distribution. We demonstrated in particular that sufficient neural statistics can be achieved using the VMIM loss function, in which case both implicit and explicit full-field inference yield the same posterior.

We created an experimental setup specifically designed to assess the effects of compression methods. Our custom \href{https://github.com/DifferentiableUniverseInitiative/sbi_lens}{\sbilens} package features a JAX-based lognormal weak lensing forward model that allows us to use the forward model for explicit full-field inference and simulate the mock data required to train the implicit model.
 It is designed for inference applications that need access to the model's derivatives. Our analysis is based on synthetic weak-lensing data with five tomographic bins, mimicking a survey like LSST Y10.

After providing an overview of the different compression strategies adopted in the literature for implicit inference strategies, we compared the impact of some of those on the final constraints on the cosmological parameters for a $w$CDM model. We also performed the classical power spectrum analysis and the full-field explicit analysis which consists of directly sampling the BHM.
We found the following results:
\begin{enumerate}
    \item The marginalized summary statistics indicate that VMIM produces better results for $\Omega_c$, $w_0$, and $\sigma_8$ in terms of agreement with the fiducial value. However, it is important to note that the results from MSE,  and MAE are not in tension with the fiducial parameters.
    Furthermore, we quantified the outcomes by examining the FoM and found that VMIM provides more precise measurements compared to MSE, and MAE.
    \item When using the VMIM to compress the original high-dimensional data, we compared the posterior obtained in the implicit inference framework with those obtained from Bayesian hierarchical modeling and the power spectrum. We demonstrate that both map-based approaches lead to a significant improvement in constraining $\Omega_c, w_0, \sigma_8$ compared to the 2-point statistics. However, $n_s$ is not more constrained with full-field inference than with two-point inference and $h, \Omega_b$ are not constrained by either and are prior-dominated.
    \item When using the VMIM to compress the original high-dimensional data the two methods, i.e., Bayesian hierarchical inference and implicit inference, lead to the same posterior distributions.
\end{enumerate}

It is important to consider the potential limitations of the current implementation and highlight particular strategies for future extensions and applications of this project. In this work, we employed a physical model based on a lognormal field, which is notably faster than N-body simulation-based methods. Although we have shown that this description accounts for additional non-Gaussian information, as evidenced by the fact that we obtain different posteriors from the full-field and power spectrum methods, it is important to note that this is a good approximation for the convergence at intermediate scales but may not be appropriate for analyzing small scales. 
Furthermore, the lognormal shift parameters are computed using the \texttt{Cosmomentum} code \citep{friedrich2018density, friedrich2020primordial}, which employs perturbation theory. As mentioned by \citet{boruah2022map}, the perturbation theory-based approach may not provide accurate results at small scales.\footnote{However, as the main objective of this project is to compare the different inference strategies, we are not very concerned with the potential implications of this approximation at this stage.} Additionally, we did not include any systematics in the current application, although previous studies demonstrated that the map-based approaches help to dramatically improve the constraints on systematics and cosmological parameters \citep{kacprzak2022deeplss}.
Hence, the natural next step will be to implement an N-body model as the physical model for the \sbilens\ package \citep{refId0}, and to improve the realism of the model by including additional systematics such as redshift uncertainties, baryonic feedback, and a more realistic intrinsic alignment model. However, regardless of the complexity of the simulations, VMIM is theoretically capable of constructing sufficient statistics, as it maximizes the mutual information $I(\bm{t},\bm{\theta})$ between the summary statistics and the parameters $\bm{\theta}$, and by definition, a statistics is sufficient if and only if $I(\bm{t},\bm{\theta}) = I(\bm{x},\bm{\theta})$. Therefore, in theory, VMIM should also generate sufficient statistics when using more realistic simulations, however, empirical confirmation is still required, which we leave for future work.

Regarding inference methodologies, we highlight that in this work we did not take into consideration the cost of the number of simulations as an axis in this comparison. Our main goal was to bring theoretical and empirical understanding of the impact of different loss functions in the asymptotic regime of a powerful enough neural compressor, and limitless number of simulations. One possible strategy to learn statistics at a low number of simulations is the IMNN, as it only requires gradients of simulations around the fiducial cosmology. Another potential strategy first explored in \cite{Charma2024} is to use transfer learning and pre-train a compressor on fairly inexpensive simulations (e.g. FastPM dark matter only simulations), and fine tune the compression model on more realistic and expensive simulations. Given that, as illustrated in this work, we know that full-field implicit inference under VMIM compression is theoretically optimal, the remaining question on the methodology side is to optimize the end-to-end simulation cost of the method.

Regarding the metrics, the FoM reported in \autoref{tab:f_o_m} was computed from a single map. Ideally, the FoM should be averaged over multiple realizations to provide more robust results. While obtaining posterior distributions for multiple observations would be straightforward using the implicit full-field inference approach specifically through NPE, as it only requires evaluating the learned NF on new observations, applying the explicit full-field inference method would be computationally intensive. The latter would require sampling the entire forward model from scratch for each new observation.
Additionally, it is worth noting that a similar FoM for VMIM and explicit inference is necessary but not sufficient to prove that the statistics are sufficient. To provide a more comprehensive comparison, additional metrics, such as contour plots, means, and standard deviations, which complement the FoM and align with the conventional set of metrics in the field, should be considered.
Note also, that in our companion paper \citep{zeghal2024simulationbasedinferencebenchmarklsst}, we use the classifier two-sample test (C2ST) to assess the similarity between posterior distributions from VMIM and explicit inference. The C2ST score reflects the probability that samples are drawn from the same distribution, with a score of 0.5 indicating similarity and 1.0 indicating total divergence. Our analysis shows a C2ST score of 0.6, suggesting that while the distributions differ somewhat, they are not completely dissimilar. This observed difference could be due to various factors, including imperfections in explicit inference, slight insufficiencies in the summary statistics, training issues in the NF, or biases in the C2ST metric.
Although achieving exact sufficiency is challenging, the near-sufficiency of VMIM makes it a strong candidate for data compression compared to other schemes.

%--------------------------------------------------------------------
%--------------------------------------------------------------------
\begin{acknowledgements}
This work was granted access to the HPC/AI resources of IDRIS under the allocations 2022-AD011013922 and 2023-AD010414029 made by GENCI. 
This research was supported by the Munich Institute for Astro-, Particle and BioPhysics (MIAPbP), which is funded by the Deutsche Forschungsgemeinschaft (DFG, German Research Foundation) under Germany´s Excellence Strategy – EXC-2094 – 390783311.  This work was supported by the
TITAN ERA Chair project (contract no. 101086741) within the Horizon Europe Framework Program of the European Commission, and the  Agence Nationale de la Recherche (ANR-22-CE31-0014-01 TOSCA). This work was supported
by the Data Intelligence Institute of Paris (diiP), and IdEx Université de Paris
(ANR-18-IDEX-0001).
\end{acknowledgements}
%--------------------------------------------------------------------
%              #########    START BIBLIO   #########
%--------------------------------------------------------------------
\bibliographystyle{aa} % style aa.bst
\bibliography{biblio} 

\begin{thebibliography}{90}
\expandafter\ifx\csname natexlab\endcsname\relax\def\natexlab#1{#1}\fi

\bibitem[{Ajani {et~al.}(2020)Ajani, Peel, Pettorino, Starck, Li, \& Liu}]{ajani2020constraining}
Ajani, V., Peel, A., Pettorino, V., {et~al.} 2020, Physical Review D, 102, 103531

\bibitem[{Ajani {et~al.}(2021)Ajani, Starck, \& Pettorino}]{ajani2021starlet}
Ajani, V., Starck, J.-L., \& Pettorino, V. 2021, Astronomy \& Astrophysics, 645, L11

\bibitem[{{Akhmetzhanova} {et~al.}(2024){Akhmetzhanova}, {Mishra-Sharma}, \& {Dvorkin}}]{Akhmetzhanova2024}
{Akhmetzhanova}, A., {Mishra-Sharma}, S., \& {Dvorkin}, C. 2024, \mnras, 527, 7459

\bibitem[{Alemi {et~al.}(2019)Alemi, Fischer, Dillon, \& Murphy}]{alemi2019deepvariationalinformationbottleneck}
Alemi, A.~A., Fischer, I., Dillon, J.~V., \& Murphy, K. 2019, Deep Variational Information Bottleneck

\bibitem[{Alsing {et~al.}(2017)Alsing, Heavens, \& Jaffe}]{alsing2017cosmological}
Alsing, J., Heavens, A., \& Jaffe, A.~H. 2017, Monthly Notices of the Royal Astronomical Society, 466, 3272

\bibitem[{Alsing \& Wandelt(2018)}]{Alsing_2018}
Alsing, J. \& Wandelt, B. 2018, Monthly Notices of the Royal Astronomical Society: Letters, 476, L60–L64

\bibitem[{Barber \& Agakov(2003)}]{barber2003information}
Barber, D. \& Agakov, F. 2003, Advances in Neural Information Processing Systems, 16

\bibitem[{Bernardo \& Smith(2001)}]{bayesiantheory}
Bernardo, J.~M. \& Smith, A. F.~M. 2001, Measurement Science and Technology, 12, 221

\bibitem[{Bingham {et~al.}(2019)Bingham, Chen, Jankowiak, Obermeyer, Pradhan, Karaletsos, Singh, Szerlip, Horsfall, \& Goodman}]{bingham2019pyro}
Bingham, E., Chen, J.~P., Jankowiak, M., {et~al.} 2019, J. Mach. Learn. Res., 20, 28:1

\bibitem[{B{\"o}hm {et~al.}(2017)B{\"o}hm, Hilbert, Greiner, \& En{\ss}lin}]{bohm2017bayesian}
B{\"o}hm, V., Hilbert, S., Greiner, M., \& En{\ss}lin, T.~A. 2017, Physical Review D, 96, 123510

\bibitem[{Boruah {et~al.}(2022)Boruah, Rozo, \& Fiedorowicz}]{boruah2022map}
Boruah, S.~S., Rozo, E., \& Fiedorowicz, P. 2022, Monthly Notices of the Royal Astronomical Society, 516, 4111

\bibitem[{Boyle {et~al.}(2021)Boyle, Uhlemann, Friedrich, Barthelemy, Codis, Bernardeau, Giocoli, \& Baldi}]{boyle2021nuw}
Boyle, A., Uhlemann, C., Friedrich, O., {et~al.} 2021, Monthly Notices of the Royal Astronomical Society, 505, 2886

\bibitem[{Campagne {et~al.}(2023)Campagne, Lanusse, Zuntz, Boucaud, Casas, Karamanis, Kirkby, Lanzieri, Peel, \& Li}]{Campagne_2023}
Campagne, J.-E., Lanusse, F., Zuntz, J., {et~al.} 2023, The Open Journal of Astrophysics, 6

\bibitem[{Charnock {et~al.}(2018)Charnock, Lavaux, \& Wandelt}]{charnock2018automatic}
Charnock, T., Lavaux, G., \& Wandelt, B.~D. 2018, Physical Review D, 97, 083004

\bibitem[{Cheng \& M{\'e}nard(2021)}]{cheng2021weak}
Cheng, S. \& M{\'e}nard, B. 2021, Monthly Notices of the Royal Astronomical Society, 507, 1012

\bibitem[{Clerkin {et~al.}(2017)Clerkin, Kirk, Manera, Lahav, Abdalla, Amara, Bacon, Chang, Gaztanaga, Hawken, {et~al.}}]{clerkin2017testing}
Clerkin, L., Kirk, D., Manera, M., {et~al.} 2017, Monthly Notices of the Royal Astronomical Society, 466, 1444

\bibitem[{Coles \& Jones(1991)}]{coles1991lognormal}
Coles, P. \& Jones, B. 1991, Monthly Notices of the Royal Astronomical Society, 248, 1

\bibitem[{Cranmer {et~al.}(2020)Cranmer, Brehmer, \& Louppe}]{cranmer2020frontier}
Cranmer, K., Brehmer, J., \& Louppe, G. 2020, Proceedings of the National Academy of Sciences, 117, 30055

\bibitem[{Cranmer {et~al.}(2015)Cranmer, Pavez, \& Louppe}]{lr1}
Cranmer, K., Pavez, J., \& Louppe, G. 2015, Approximating Likelihood Ratios with Calibrated Discriminative Classifiers

\bibitem[{Dai \& Seljak(2024)}]{Dai2024}
Dai, B. \& Seljak, U. 2024, Proceedings of the National Academy of Sciences, 121, e2309624121

\bibitem[{Dinh {et~al.}(2017)Dinh, Sohl-Dickstein, \& Bengio}]{realnvp}
Dinh, L., Sohl-Dickstein, J., \& Bengio, S. 2017, Density estimation using Real NVP

\bibitem[{Elfwing {et~al.}(2017)Elfwing, Uchibe, \& Doya}]{silu}
Elfwing, S., Uchibe, E., \& Doya, K. 2017, Sigmoid-Weighted Linear Units for Neural Network Function Approximation in Reinforcement Learning

\bibitem[{{Fiedorowicz} {et~al.}(2022){Fiedorowicz}, {Rozo}, \& {Boruah}}]{Karma2022}
{Fiedorowicz}, P., {Rozo}, E., \& {Boruah}, S.~S. 2022, arXiv e-prints, arXiv:2210.12280

\bibitem[{Fiedorowicz {et~al.}(2022)Fiedorowicz, Rozo, Boruah, Chang, \& Gatti}]{Fiedorowicz_2022}
Fiedorowicz, P., Rozo, E., Boruah, S.~S., Chang, C., \& Gatti, M. 2022, Monthly Notices of the Royal Astronomical Society, 512, 73–85

\bibitem[{Fluri {et~al.}(2019)Fluri, Kacprzak, Lucchi, Refregier, Amara, Hofmann, \& Schneider}]{fluri2019cosmological}
Fluri, J., Kacprzak, T., Lucchi, A., {et~al.} 2019, Physical Review D, 100, 063514

\bibitem[{Fluri {et~al.}(2022)Fluri, Kacprzak, Lucchi, Schneider, Refregier, \& Hofmann}]{fluri2022full}
Fluri, J., Kacprzak, T., Lucchi, A., {et~al.} 2022, Physical Review D, 105, 083518

\bibitem[{Fluri {et~al.}(2018)Fluri, Kacprzak, Refregier, Amara, Lucchi, \& Hofmann}]{fluri2018cosmological}
Fluri, J., Kacprzak, T., Refregier, A., {et~al.} 2018, Physical Review D, 98, 123518

\bibitem[{Fluri {et~al.}(2021)Fluri, Kacprzak, Refregier, Lucchi, \& Hofmann}]{fluri2021cosmological}
Fluri, J., Kacprzak, T., Refregier, A., Lucchi, A., \& Hofmann, T. 2021, Physical Review D, 104, 123526

\bibitem[{Friedrich {et~al.}(2018)Friedrich, Gruen, DeRose, Kirk, Krause, McClintock, Rykoff, Seitz, Wechsler, Bernstein, {et~al.}}]{friedrich2018density}
Friedrich, O., Gruen, D., DeRose, J., {et~al.} 2018, Physical Review D, 98, 023508

\bibitem[{Friedrich {et~al.}(2020)Friedrich, Uhlemann, Villaescusa-Navarro, Baldauf, Manera, \& Nishimichi}]{friedrich2020primordial}
Friedrich, O., Uhlemann, C., Villaescusa-Navarro, F., {et~al.} 2020, Monthly Notices of the Royal Astronomical Society, 498, 464

\bibitem[{Gatti {et~al.}(2021)Gatti, Jain, Chang, Raveri, Z{\"u}rcher, Secco, Whiteway, Jeffrey, Doux, Kacprzak, {et~al.}}]{gatti2021dark}
Gatti, M., Jain, B., Chang, C., {et~al.} 2021, arXiv preprint arXiv:2110.10141

\bibitem[{Greenberg {et~al.}(2019)Greenberg, Nonnenmacher, \& Macke}]{npe3}
Greenberg, D.~S., Nonnenmacher, M., \& Macke, J.~H. 2019, Automatic Posterior Transformation for Likelihood-Free Inference

\bibitem[{{Gupta} {et~al.}(2018){Gupta}, {Matilla}, {Hsu}, {et~al.}}]{2018PhRvD..97j3515G}
{Gupta}, A., {Matilla}, J. M.~Z., {Hsu}, D., {et~al.} 2018, 97, 103515

\bibitem[{Halder {et~al.}(2021)Halder, Friedrich, Seitz, \& Varga}]{halder2021integrated}
Halder, A., Friedrich, O., Seitz, S., \& Varga, T.~N. 2021, Monthly Notices of the Royal Astronomical Society, 506, 2780

\bibitem[{Harnois-D{\'e}raps {et~al.}(2021)Harnois-D{\'e}raps, Martinet, \& Reischke}]{harnois2021cosmic}
Harnois-D{\'e}raps, J., Martinet, N., \& Reischke, R. 2021, Monthly Notices of the Royal Astronomical Society

\bibitem[{He {et~al.}(2016)He, Zhang, Ren, \& Sun}]{he2016deep}
He, K., Zhang, X., Ren, S., \& Sun, J. 2016, in Proceedings of the IEEE conference on computer vision and pattern recognition, 770--778

\bibitem[{Heavens {et~al.}(2000)Heavens, Jimenez, \& Lahav}]{heavens2000massive}
Heavens, A.~F., Jimenez, R., \& Lahav, O. 2000, Monthly Notices of the Royal Astronomical Society, 317, 965

\bibitem[{Hennigan {et~al.}(2020)Hennigan, Cai, Norman, Martens, \& Babuschkin}]{haiku2020github}
Hennigan, T., Cai, T., Norman, T., Martens, L., \& Babuschkin, I. 2020, {H}aiku: {S}onnet for {JAX}

\bibitem[{Hilbert {et~al.}(2011)Hilbert, Hartlap, \& Schneider}]{hilbert2011cosmic}
Hilbert, S., Hartlap, J., \& Schneider, P. 2011, Astronomy \& Astrophysics, 536, A85

\bibitem[{Hoffman {et~al.}(2014)Hoffman, Gelman, {et~al.}}]{hoffman2014no}
Hoffman, M.~D., Gelman, A., {et~al.} 2014, J. Mach. Learn. Res., 15, 1593

\bibitem[{Ivezi{\'c} {et~al.}(2019)Ivezi{\'c}, Kahn, Tyson, Abel, Acosta, Allsman, Alonso, AlSayyad, Anderson, Andrew, {et~al.}}]{ivezic2019lsst}
Ivezi{\'c}, {\v{Z}}., Kahn, S.~M., Tyson, J.~A., {et~al.} 2019, The Astrophysical Journal, 873, 111

\bibitem[{Izbicki {et~al.}(2014)Izbicki, Lee, \& Schafer}]{lr2}
Izbicki, R., Lee, A.~B., \& Schafer, C.~M. 2014

\bibitem[{Jaynes(2003)}]{jaynes03}
Jaynes, E.~T. 2003, Probability theory: The logic of science (Cambridge: Cambridge University Press)

\bibitem[{Jeffrey {et~al.}(2021)Jeffrey, Alsing, \& Lanusse}]{jeffrey2021likelihood}
Jeffrey, N., Alsing, J., \& Lanusse, F. 2021, Monthly Notices of the Royal Astronomical Society, 501, 954

\bibitem[{{Jeffrey} {et~al.}(2024){Jeffrey}, {Whiteway}, {Gatti}, {Williamson}, {Alsing}, {Porredon}, {Prat}, {Doux}, {Jain}, {Chang}, {Cheng}, {Kacprzak}, {Lemos}, {Alarcon}, {Amon}, {Bechtol}, {Becker}, {Bernstein}, {Campos}, {Carnero Rosell}, {Chen}, {Choi}, {DeRose}, {Drlica-Wagner}, {Eckert}, {Everett}, {Fert{\'e}}, {Gruen}, {Gruendl}, {Herner}, {Jarvis}, {McCullough}, {Myles}, {Navarro-Alsina}, {Pandey}, {Raveri}, {Rollins}, {Rykoff}, {S{\'a}nchez}, {Secco}, {Sevilla-Noarbe}, {Sheldon}, {Shin}, {Troxel}, {Tutusaus}, {Varga}, {Yanny}, {Yin}, {Zuntz}, {Aguena}, {Allam}, {Alves}, {Bacon}, {Bocquet}, {Brooks}, {da Costa}, {Davis}, {De Vicente}, {Desai}, {Diehl}, {Ferrero}, {Frieman}, {Garc{\'\i}a-Bellido}, {Gaztanaga}, {Giannini}, {Gutierrez}, {Hinton}, {Hollowood}, {Honscheid}, {Huterer}, {James}, {Lahav}, {Lee}, {Marshall}, {Mena-Fern{\'a}ndez}, {Miquel}, {Pieres}, {Plazas Malag{\'o}n}, {Roodman}, {Sako}, {Sanchez}, {Sanchez Cid}, {Smith}, {Suchyta}, {Swanson}, {Tarle}, {Tucker}, {Weaverdyck}, {Weller}, {Wiseman}, \& {Yamamoto}}]{Jeffrey2024}
{Jeffrey}, N., {Whiteway}, L., {Gatti}, M., {et~al.} 2024, arXiv e-prints, arXiv:2403.02314

\bibitem[{{Junzhe Zhou} {et~al.}(2023){Junzhe Zhou}, {Li}, {Dodelson}, \& {Mandelbaum}}]{Zhou2023}
{Junzhe Zhou}, A., {Li}, X., {Dodelson}, S., \& {Mandelbaum}, R. 2023, arXiv e-prints, arXiv:2312.08934

\bibitem[{Kacprzak \& Fluri(2022)}]{kacprzak2022deeplss}
Kacprzak, T. \& Fluri, J. 2022, Physical Review X, 12, 031029

\bibitem[{Kacprzak {et~al.}(2016)Kacprzak, Kirk, Friedrich, Amara, Refregier, Marian, Dietrich, Suchyta, Aleksi{\'c}, Bacon, {et~al.}}]{kacprzak2016cosmology}
Kacprzak, T., Kirk, D., Friedrich, O., {et~al.} 2016, Monthly Notices of the Royal Astronomical Society, 463, 3653

\bibitem[{Klypin {et~al.}(2018)Klypin, Prada, Betancort-Rijo, \& Albareti}]{klypin2018density}
Klypin, A., Prada, F., Betancort-Rijo, J., \& Albareti, F.~D. 2018, Monthly Notices of the Royal Astronomical Society, 481, 4588

\bibitem[{Kratochvil {et~al.}(2012)Kratochvil, Lim, Wang, Haiman, May, \& Huffenberger}]{kratochvil2012probing}
Kratochvil, J.~M., Lim, E.~A., Wang, S., {et~al.} 2012, Physical Review D, 85, 103513

\bibitem[{Kullback \& Leibler(1951)}]{kullback1951information}
Kullback, S. \& Leibler, R.~A. 1951, The annals of mathematical statistics, 22, 79

\bibitem[{{Lanzieri, Denise} {et~al.}(2023){Lanzieri, Denise}, {Lanusse, Fran\c{c}ois}, {Modi, Chirag}, {Horowitz, Benjamin}, {Harnois-D\'eraps, Joachim}, {Starck, Jean-Luc}, \& {The LSST Dark Energy Science Collaboration (LSST DESC)}}]{refId0}
{Lanzieri, Denise}, {Lanusse, Fran\c{c}ois}, {Modi, Chirag}, {et~al.} 2023, A\&A, 679, A61

\bibitem[{Laureijs {et~al.}(2011)Laureijs, Amiaux, Arduini, Augueres, Brinchmann, Cole, Cropper, Dabin, Duvet, Ealet, {et~al.}}]{laureijs2011euclid}
Laureijs, R., Amiaux, J., Arduini, S., {et~al.} 2011, arXiv preprint arXiv:1110.3193

\bibitem[{Lin \& Kilbinger(2015)}]{lin2015new}
Lin, C.-A. \& Kilbinger, M. 2015, Astronomy \& Astrophysics, 583, A70

\bibitem[{Liu \& Madhavacheril(2019)}]{liu2019constraining}
Liu, J. \& Madhavacheril, M.~S. 2019, Physical Review D, 99, 083508

\bibitem[{Liu {et~al.}(2015{\natexlab{a}})Liu, Petri, Haiman, Hui, Kratochvil, \& May}]{liu2015cosmology}
Liu, J., Petri, A., Haiman, Z., {et~al.} 2015{\natexlab{a}}, Physical Review D, 91, 063507

\bibitem[{Liu {et~al.}(2015{\natexlab{b}})Liu, Pan, Li, Shan, Wang, Fu, Fan, Kneib, Leauthaud, Van~Waerbeke, {et~al.}}]{liu2015cosmological}
Liu, X., Pan, C., Li, R., {et~al.} 2015{\natexlab{b}}, Monthly Notices of the Royal Astronomical Society, 450, 2888

\bibitem[{Lu {et~al.}(2023)Lu, Haiman, \& Li}]{lu2023cosmological}
Lu, T., Haiman, Z., \& Li, X. 2023, arXiv preprint arXiv:2301.01354

\bibitem[{Lu {et~al.}(2022)Lu, Haiman, \& Zorrilla~Matilla}]{lu2022simultaneously}
Lu, T., Haiman, Z., \& Zorrilla~Matilla, J.~M. 2022, Monthly Notices of the Royal Astronomical Society, 511, 1518

\bibitem[{Lueckmann {et~al.}(2018)Lueckmann, Bassetto, Karaletsos, \& Macke}]{nle2}
Lueckmann, J.-M., Bassetto, G., Karaletsos, T., \& Macke, J.~H. 2018

\bibitem[{Lueckmann {et~al.}(2017)Lueckmann, Goncalves, Bassetto, Öcal, Nonnenmacher, \& Macke}]{npe2}
Lueckmann, J.-M., Goncalves, P.~J., Bassetto, G., {et~al.} 2017, Flexible statistical inference for mechanistic models of neural dynamics

\bibitem[{Makinen {et~al.}(2021)Makinen, Charnock, Alsing, \& Wandelt}]{Makinen_2021}
Makinen, T.~L., Charnock, T., Alsing, J., \& Wandelt, B.~D. 2021, Journal of Cosmology and Astroparticle Physics, 2021, 049

\bibitem[{Makinen {et~al.}(2022)Makinen, Charnock, Lemos, Porqueres, Heavens, \& Wandelt}]{Makinen_2022}
Makinen, T.~L., Charnock, T., Lemos, P., {et~al.} 2022, The Open Journal of Astrophysics, 5

\bibitem[{Mandelbaum {et~al.}(2018)Mandelbaum, Eifler, Hlo{\v{z}}ek, Collett, Gawiser, Scolnic, Alonso, Awan, Biswas, Blazek, {et~al.}}]{mandelbaum2018lsst}
Mandelbaum, R., Eifler, T., Hlo{\v{z}}ek, R., {et~al.} 2018, arXiv preprint arXiv:1809.01669

\bibitem[{Martinet {et~al.}(2018)Martinet, Schneider, Hildebrandt, Shan, Asgari, Dietrich, Harnois-D{\'e}raps, Erben, Grado, Heymans, {et~al.}}]{martinet2018kids}
Martinet, N., Schneider, P., Hildebrandt, H., {et~al.} 2018, Monthly Notices of the Royal Astronomical Society, 474, 712

\bibitem[{Matilla {et~al.}(2020)Matilla, Sharma, Hsu, \& Haiman}]{PhysRevD.102.123506}
Matilla, J. M.~Z., Sharma, M., Hsu, D., \& Haiman, Z. 2020, Phys. Rev. D, 102, 123506

\bibitem[{Papamakarios \& Murray(2018)}]{npe1}
Papamakarios, G. \& Murray, I. 2018, Fast $\epsilon$-free Inference of Simulation Models with Bayesian Conditional Density Estimation

\bibitem[{Papamakarios {et~al.}(2018)Papamakarios, Sterratt, \& Murray}]{nle1}
Papamakarios, G., Sterratt, D.~C., \& Murray, I. 2018, Sequential Neural Likelihood: Fast Likelihood-free Inference with Autoregressive Flows

\bibitem[{Peel {et~al.}(2017)Peel, Lin, Lanusse, Leonard, Starck, \& Kilbinger}]{peel2017cosmological}
Peel, A., Lin, C.-A., Lanusse, F., {et~al.} 2017, Astronomy \& Astrophysics, 599, A79

\bibitem[{{Petri}(2016)}]{2016A&C....17...73P}
{Petri}, A. 2016, Astronomy and Computing, 17, 73

\bibitem[{Petri {et~al.}(2013)Petri, Haiman, Hui, May, \& Kratochvil}]{petri2013cosmology}
Petri, A., Haiman, Z., Hui, L., May, M., \& Kratochvil, J.~M. 2013, Physical Review D, 88, 123002

\bibitem[{Phan {et~al.}(2019)Phan, Pradhan, \& Jankowiak}]{phan2019composable}
Phan, D., Pradhan, N., \& Jankowiak, M. 2019, arXiv preprint arXiv:1912.11554

\bibitem[{Porqueres {et~al.}(2021)Porqueres, Heavens, Mortlock, \& Lavaux}]{porqueres2021bayesian}
Porqueres, N., Heavens, A., Mortlock, D., \& Lavaux, G. 2021, Monthly Notices of the Royal Astronomical Society, 502, 3035

\bibitem[{Porqueres {et~al.}(2023)Porqueres, Heavens, Mortlock, Lavaux, \& Makinen}]{porqueres2023field}
Porqueres, N., Heavens, A., Mortlock, D., Lavaux, G., \& Makinen, T.~L. 2023, arXiv preprint arXiv:2304.04785

\bibitem[{Ribli {et~al.}(2019)Ribli, Pataki, Zorrilla~Matilla, Hsu, Haiman, \& Csabai}]{ribli2019weak}
Ribli, D., Pataki, B.~{\'A}., Zorrilla~Matilla, J.~M., {et~al.} 2019, Monthly Notices of the Royal Astronomical Society, 490, 1843

\bibitem[{Ribli {et~al.}(2018)Ribli, Ármin Pataki, \& Csabai}]{ribli2018improved}
Ribli, D., Ármin Pataki, B., \& Csabai, I. 2018, An improved cosmological parameter inference scheme motivated by deep learning

\bibitem[{Rizzato {et~al.}(2019)Rizzato, Benabed, Bernardeau, \& Lacasa}]{rizzato2019tomographic}
Rizzato, M., Benabed, K., Bernardeau, F., \& Lacasa, F. 2019, Monthly Notices of the Royal Astronomical Society, 490, 4688

\bibitem[{Semboloni {et~al.}(2011)Semboloni, Schrabback, van Waerbeke, Vafaei, Hartlap, \& Hilbert}]{semboloni2011weak}
Semboloni, E., Schrabback, T., van Waerbeke, L., {et~al.} 2011, Monthly Notices of the Royal Astronomical Society, 410, 143

\bibitem[{Shan {et~al.}(2018)Shan, Liu, Hildebrandt, Pan, Martinet, Fan, Schneider, Asgari, Harnois-D{\'e}raps, Hoekstra, {et~al.}}]{shan2018kids}
Shan, H., Liu, X., Hildebrandt, H., {et~al.} 2018, Monthly Notices of the Royal Astronomical Society, 474, 1116

\bibitem[{{Sharma} {et~al.}(2024){Sharma}, {Dai}, \& {Seljak}}]{Charma2024}
{Sharma}, D., {Dai}, B., \& {Seljak}, U. 2024, arXiv e-prints, arXiv:2403.03490

\bibitem[{Smail {et~al.}(1995)Smail, Hogg, Yan, \& Cohen}]{smail1995deep}
Smail, I., Hogg, D.~W., Yan, L., \& Cohen, J.~G. 1995, The Astrophysical Journal, 449, L105

\bibitem[{Spergel {et~al.}(2015)Spergel, Gehrels, Baltay, Bennett, Breckinridge, Donahue, Dressler, Gaudi, Greene, Guyon, {et~al.}}]{spergel2015wide}
Spergel, D., Gehrels, N., Baltay, C., {et~al.} 2015, arXiv preprint arXiv:1503.03757

\bibitem[{Takada \& Jain(2004)}]{takada2004cosmological}
Takada, M. \& Jain, B. 2004, Monthly Notices of the Royal Astronomical Society, 348, 897

\bibitem[{Thomas {et~al.}(2016)Thomas, Dutta, Corander, Kaski, \& Gutmann}]{lr3}
Thomas, O., Dutta, R., Corander, J., Kaski, S., \& Gutmann, M.~U. 2016, Likelihood-free inference by ratio estimation

\bibitem[{Tishby {et~al.}(2000)Tishby, Pereira, \& Bialek}]{tishby2000informationbottleneckmethod}
Tishby, N., Pereira, F.~C., \& Bialek, W. 2000, The information bottleneck method

\bibitem[{Uhlemann {et~al.}(2020)Uhlemann, Friedrich, Villaescusa-Navarro, Banerjee, \& Codis}]{uhlemann2020fisher}
Uhlemann, C., Friedrich, O., Villaescusa-Navarro, F., Banerjee, A., \& Codis, S. 2020, Monthly Notices of the Royal Astronomical Society, 495, 4006

\bibitem[{Xavier {et~al.}(2016)Xavier, Abdalla, \& Joachimi}]{xavier2016improving}
Xavier, H.~S., Abdalla, F.~B., \& Joachimi, B. 2016, Monthly Notices of the Royal Astronomical Society, 459, 3693

\bibitem[{Zeghal {et~al.}(2024)Zeghal, Lanzieri, Lanusse, Boucaud, Louppe, Aubourg, Bayer, \& Collaboration}]{zeghal2024simulationbasedinferencebenchmarklsst}
Zeghal, J., Lanzieri, D., Lanusse, F., {et~al.} 2024, Simulation-Based Inference Benchmark for LSST Weak Lensing Cosmology

\bibitem[{Zhang {et~al.}(2022)Zhang, Chang, Larsen, Secco, Zuntz, \& Collaboration}]{zhang2022transitioning}
Zhang, Z., Chang, C., Larsen, P., {et~al.} 2022, Monthly Notices of the Royal Astronomical Society, 514, 2181

\bibitem[{Z{\"u}rcher {et~al.}(2022)Z{\"u}rcher, Fluri, Sgier, Kacprzak, Gatti, Doux, Whiteway, R{\'e}fr{\'e}gier, Chang, Jeffrey, {et~al.}}]{zurcher2022dark}
Z{\"u}rcher, D., Fluri, J., Sgier, R., {et~al.} 2022, Monthly Notices of the Royal Astronomical Society, 511, 2075

\end{thebibliography}
%--------------------------------------------------------------------
%--------------------------------------------------------------------

%--------------------------------------------------------------------
%              #########    START APPENDIX   #########
%--------------------------------------------------------------------

\begin{appendix}
\section{Mean Squared Error (MSE)}\label{Sec:appendix_Mean Square Error}
In this section, as well as the next one, we derive established results concerning the regression loss functions and what they actually learn. Additional details can be found in \cite{jaynes03}.

Here, we demonstrate that minimizing the $L_2$ norm is equivalent to training the model to estimate the mean of the posterior distribution. Namely:
\begin{equation}\label{Eq:mean_mse}
\left \langle \bm {\theta} \right \rangle_{p(\bm {\theta}|\bm{x})}=\operatorname*{argmin}_{F(\bm{x})}\mathbb{E}_{p(\bm {\theta},\bm {x})}[\left\Vert \bm {\theta}-F(\bm{x})
 \right \Vert_{2}^{2}],
\end{equation}
where the posterior mean $\left \langle \bm {\theta} \right \rangle_{p(\bm {\theta}|\bm{x})}$, is calculated as follows:
\begin{equation}\label{Eq:mean_pos}
     \left \langle \bm {\theta} \right \rangle_{p(\bm {\theta}|\bm{x})}= \mathbb{E}_{p(\bm {\theta}|\bm {x})}[\bm {\theta}].
\end{equation}
To demonstrate this statement, we need to minimize the expected value of the $L_2$ norm with respect to $F(\bm{x})$. Let us consider its derivative:
\begin{align}\label{Eq:moment_1}
   & \frac{\partial}{\partial F(\bm{x}) }  \mathbb{E}_{p(\bm {\theta}|\bm{x}))}[(\bm {\theta}-F(\bm{x}))^2] =  \\
    &
    \frac{\partial}{\partial F(\bm{x}) }  \mathbb{E}_{p(\bm {\theta}|\bm {x})}[\bm {\theta}^2+F(\bm{x})^2-2\bm {\theta}F(\bm{x})] = \nonumber \\
    &
    \frac{\partial}{\partial F(\bm{x}) }  [
    \mathbb{E}_{p(\bm {\theta}|\bm {x})}[\bm {\theta}^2]+F(\bm{x})^2-2F(\bm{x})\mathbb{E}_{p(\bm {\theta}|\bm {x})}[\bm {\theta}]]= \nonumber \\
    &2F(\bm{x})-2 \mathbb{E}_{p(\bm {\theta}|\bm {x})}[\bm {\theta}]. \nonumber
\end{align}
Setting it equal to zero, we obtain the critical value:
\begin{equation}\label{Eq:minimum_mean}
    F(\bm{x})= \mathbb{E}_{p(\bm {\theta}|\bm {x})}[\bm {\theta}]. 
\end{equation}
Considering the second-order derivative:
\begin{equation}
   \frac{\partial^2}{\partial^2 F(\bm{x})}  \mathbb{E}_{p(\bm {\theta}|\bm {x})}[(\bm {\theta}-F(\bm{x}))^2]=2>0, 
\end{equation}
we can assert that this critical value is also a minimum. \\
From \autoref{Eq:minimum_mean} and \autoref{Eq:mean_pos}, we obtain \autoref{Eq:mean_mse}.
%--------------------------------------------------------------------
\section{Mean Absolute Error (MAE)}\label{Sec:appendix_Maximum Absolute Error}
In this section, we demonstrate that minimizing the $L_1$ norm is equivalent to training the model to estimate the median of the posterior distribution.
Namely:
\begin{equation}\label{Eq:median_mae}
 \bm {\theta}^M_{p(\bm {\theta}|\bm{x})}=\operatorname*{argmin}_{F(\bm{x})}\mathbb{E}_{p(\bm {\theta},\bm {x})}[| \bm {\theta}-F(\bm{x})|].
\end{equation}
By definition, the median of a one-dimensional\footnote{For notational simplicity, we demonstrate this statement in one dimension; the generalization to N-dimensions is straightforward.} probability density function $p(x)$ is a real number $m$ that satisfies:
\begin{equation}\label{Eq:definition_median}
\int_{\infty}^{m} p(x)dx=\int_{m}^{\infty}p(x)dx=\frac{1}{2}.
\end{equation}
The expectation value of the mean absolute error is defined as:
\begin{equation}
    \mathbb{E}_{p(x)}[|x-m|]= \int_{\infty}^{\infty}p(x)|x-m|dx  
\end{equation}
which can be decomposed as
\begin{equation}
        \int_{\infty}^{m}p(x)|x-m|dx +\int_{m}^{\infty}p(x)|x-m|dx .
\end{equation}
To minimize this function with respect to $m$, we need to compute its derivative:
\begin{equation}\label{Eq:absolute_median}
    \frac{d\mathbb{E}[|x-m|]}{dm}=
    \frac{d}{dm}\int_{\infty}^{m}p(x)|x-m|dx +\frac{d}{dm}\int_{m}^{\infty}p(x)|x-m|dx. 
\end{equation}
Considering that $|x-m|=(x-m)$ for $m\le x$ and $|x-m|=(m-x)$ $m\ge x$, 
we can write \autoref{Eq:absolute_median} as:
\begin{equation}
    \frac{d\mathbb{E}[|x-m|]}{dm}=
    \frac{d}{dm}\int_{\infty}^{m}p(x)(m-x)dx +\frac{d}{dm}\int_{m}^{\infty}p(x)(x-m)dx .
\end{equation}
Using the Leibniz integral rule, we get:
\begin{align}
    &\frac{d\mathbb{E}[|x-m|]}{dm}= \\
    &
    p(x)(m-x)\frac{dm}{dm}+\int_{\infty}^{m}\frac{\partial}{\partial m}[p(x)(m-x)]dx  \nonumber \\
    & + p(x)(x-m)\frac{dm}{dm}+\int_{m}^{\infty}\frac{\partial}{\partial m}[p(x)(x-m)]dx \nonumber .
\end{align}
Setting the derivative to zero, we obtain:
\begin{equation}
    \frac{d\mathbb{E}[|x-m|]}{dm}= \int_{\infty}^{m} p(x)dx-\int_{m}^{\infty}p(x)dx =0.
\end{equation}
Thus,
\begin{equation}
\int_{\infty}^{m} p(x)dx=\int_{m}^{\infty}p(x)dx .
\end{equation}
Considering that
\begin{equation}
\int_{\infty}^{m} p(x)dx+\int_{m}^{\infty}p(x)dx=1,
\end{equation}
we obtain \autoref{Eq:definition_median}.

\section{Gaussian negative log-likelihood (GNLL)}\label{Sec:appendix_gnll}
In this section, we demonstrate that the summary statistics obtained from the GNLL minimization corresponds to the mean of the posterior exactly as the minimization of the MSE loss function. 

The minimization of the GNLL corresponds to the minimization of the forward Kullback Leiber Divergence (DKL): 
     \begin{align}
        \hat{\varphi} &=  \arg \min_{\varphi} D_{KL}(p(x) \: || \: q_{\varphi}(x)) \\
        & = \arg \min_{\varphi} \mathbb{E}_{p(x)}\Big[ \log\left(\frac{p(x)}{q_{\varphi}(x)}\right) \Big]   \nonumber\\
        \begin{split}
        & = \arg \min_{\varphi} \underbrace{\mathbb{E}_{p(x)}\left[ \log\left(p(x)\right) \right]}_{\text{constant w.r.t }\varphi} - \mathbb{E}_{p(x)}\left[ \log\left(q_{\varphi}(x)\right) \right] 
        \end{split} \nonumber \\
        & =  \arg \min_{\varphi} - \mathbb{E}_{p(x)}\left[ \log\left(q_{\varphi}( x)\right) \right] \nonumber
    \end{align}

where we let $q$ be a Gaussian distribution 
\begin{equation}
    q(x) = \frac{1}{\sqrt{2\pi\sigma_q^2}} \exp\left(-\frac{(x - \mu_q)^2}{2\sigma_q^2}\right),
\end{equation}
and $\varphi = (\mu_q, \sigma_q)$ be the mean and covariance of this distribution that we aim to optimize to minimize this forward DKL. 

Replacing $q$ by the Gaussian distribution in the forward DKL expression yields 
\begin{align}
D_{KL}(p || q) &= \int p(x) \log p(x) \, dx + \int p(x) \log \left(\sqrt{2\pi\sigma_q^2} \right) \, dx \nonumber \\
&\qquad + \int p(x) \frac{(x - \mu_q)^2}{2\sigma_q^2} \, dx. \nonumber
\end{align}

The first term of this integral is independent of the parameters $\varphi$ and corresponds to the entropy $H(p)$. The second term simplifies to $\log( \sqrt{2\pi\sigma_q^2})$, as the integral of $p(x)$ over its domain is $1$. The third term necessitates a bit more work: 
    \begin{align}
        & \int p(x) \frac{(x - \mu_q)^2}{2\sigma_q^2} \, dx  \\
        &=\frac{1}{2\sigma_q^2} \int p(x) ( (x - \mathbb{E}_p[X]) + (\mathbb{E}_p[X] - \mu_q) )^2 \, dx\\
        &= \frac{1}{2\sigma_q^2}\int p(x) (x - \mathbb{E}_p[X])^2 \, dx \\
         & \qquad+ \frac{1}{\sigma_q^2}(\mathbb{E}_p[X] - \mu_q) \int p(x) (x - \mathbb{E}_p[X]) \, dx \\
         & \qquad+ \frac{1}{2\sigma_q^2} (\mathbb{E}_p[X] - \mu_q)^2 \int p(x) \, dx, 
    \end{align}
and by definition of the expected value 
\begin{align}
    \int p(x) (x - \mathbb{E}_p[X]) \, dx &= - \mathbb{E}_p[X] + \int xp(x) \, dx \nonumber\\
     &= - \mathbb{E}_p[X] + \mathbb{E}_p[X] \, dx \nonumber\\
     &=0 \nonumber,
\end{align} 
the middle term vanishes yielding to 
 \begin{align}
    & \frac{1}{2\sigma_q^2} \left[ \int p(x) (x - \mathbb{E}_p[X])^2 \, dx + (\mathbb{E}_p[X] - \mu_q)^2 \right] \\
     &= \frac{1}{2\sigma_q^2} \left[ \text{Var}_p[X] + (\mathbb{E}_p[X] - \mu_q)^2 \right].
 \end{align}
 Finally, putting all this together we have 
 \begin{equation}
     D_{KL}(p || q) = H(p) + \log \left(\sqrt{2\pi\sigma_q^2} \right) + \frac{1}{2\sigma_q^2} \left[ \text{Var}_p[X] + (\mathbb{E}_p[X] - \mu_q)^2 \right].
 \end{equation}

Let us now find the minimum of this DKL. Computing the derivatives 
\begin{align}
    \frac{\partial  D_{KL}(p || q)}{ \partial \mu_q} &= \frac{1}{\sigma_q^2}\left(\mathbb{E}_p[X] - \mu_q \right)\\
     \frac{\partial  D_{KL}(p || q)}{ \partial \sigma_q} &= \frac{1}{\sigma_q^2} - \frac{1}{\sigma_q^3}\left[\text{Var}_p[X] + (\mathbb{E}_p[X] - \mu_q)^2\right]
\end{align}
and setting it to zero, we can derive: 
\begin{align}
\begin{cases}
 \mu_q &= \mathbb{E}_p[X]\\
\sigma_q &= \text{Var}_p[X]
\end{cases}\label{eq:critical}
\end{align}
which is the only solution of these two equations system. To confirm that this point minimizes the DKL we derive the Hessian
\begin{align}
    H(\mu_q, \sigma_q) = \begin{pmatrix} 
\frac{1}{\sigma_q^2} & -\frac{2(\mathbb{E}_p[X] - \mu_q)}{\sigma_q^3} \\[10pt]
-\frac{2(\mathbb{E}_p[X] - \mu_q)}{\sigma_q^3} & -\frac{1}{\sigma_q^2} + \frac{3(\text{Var}_p[X] + (\mathbb{E}_p[X] - \mu_q)^2)}{\sigma_q^4} 
\end{pmatrix},
\end{align}
evaluating it on the critical point we have 
\begin{align}
    H(\mathbb{E}_p[X], \text{Var}_p[X]) = \begin{pmatrix} 
\frac{1}{\text{Var}_p[X]} & 0 \\[10pt]
0 & \frac{2}{\text{Var}_p[X]} 
\end{pmatrix}.
\end{align}
To check if this symmetric matrix is positive definite we can derive
\begin{equation}
    x^T H x = \frac{x_1^2}{\text{Var}_p[X]} + \frac{2 x_2^2}{\text{Var}_p[X]},
\end{equation}
which is strictly positive for all $x = (x_1, x_2) \in \mathbb{R}^2 \backslash \{0\}$, and according to the definition this matrix is positive definite.

Hence, the critical point (\autoref{eq:critical}) corresponds to a local minimum and because it is the unique solution of the system, it is the unique minimizer of the DKL. 

%--------------------------------------------------------------------
%              #########    END APPENDIX   #########
%--------------------------------------------------------------------
%--------------------------------------------------------------------
% ############# PLOT CONVERGENCE AUTO-CROSS POWER SPECTRA  #############
%--------------------------------------------------------------------
\onecolumn
\section{Validation of \sbilens's forward model}
\begin{figure*}[h]
    \centering
    \includegraphics[width=\textwidth]{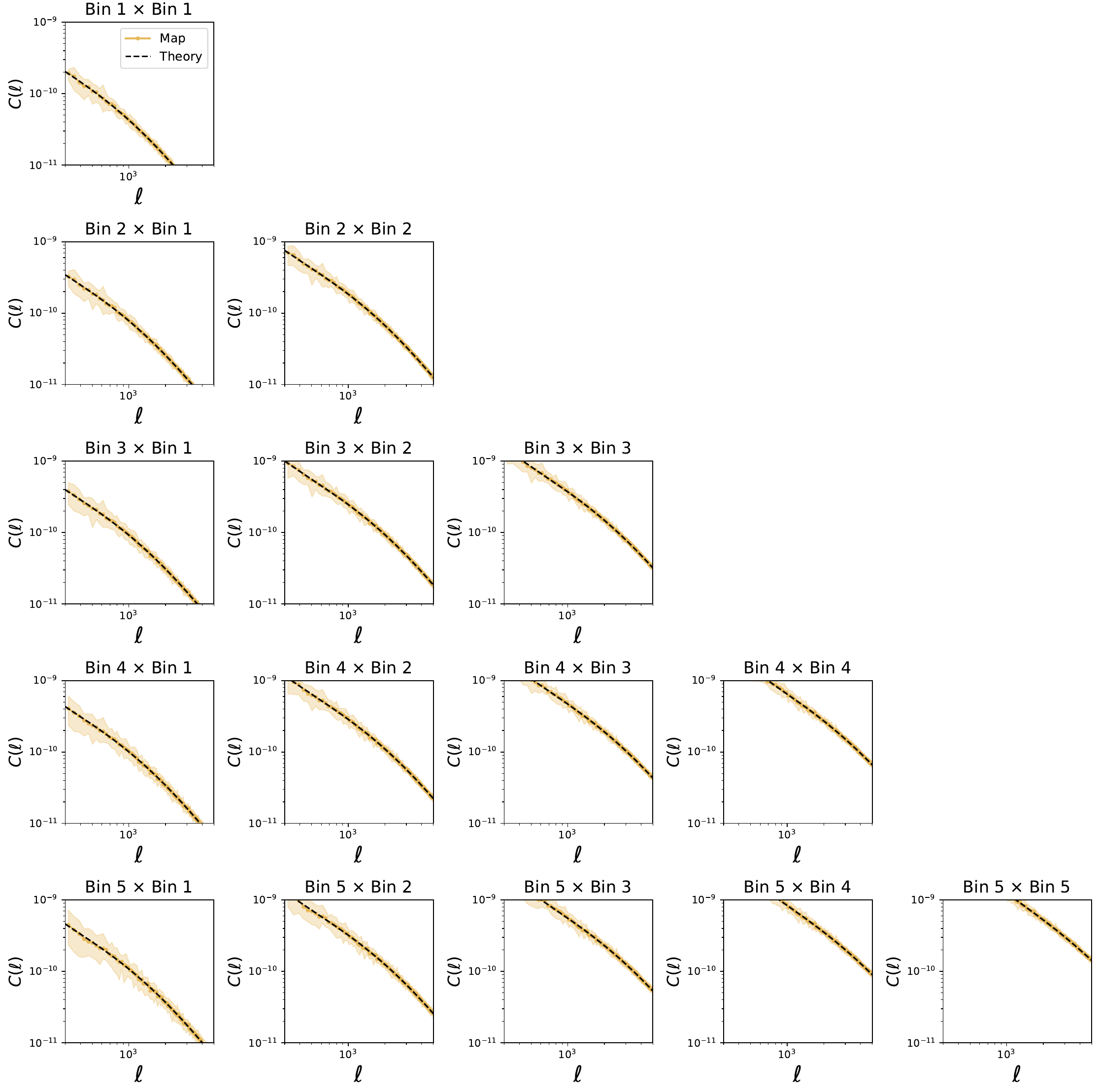}
    \caption{
    Convergence power spectra for different tomographic bin combinations. The solid yellow line shows the measurement from 20 simulated maps using the survey setting described in \autoref{Sec:the SBILens framework}, while the black dashed line shows the theoretical predictions computed using jax-cosmo. In this figure, the shaded regions represent the standard deviation from 20 independent map realizations.
    }
     \label{fig:psconvergence_maps}
\end{figure*}

\end{appendix}

\end{document}